\documentclass [11pt]{article}
\usepackage{amsfonts}
\usepackage{amsmath}
\newtheorem{theorem}{Theorem}[section]

\newtheorem{lemma}{Lemma}

\begin{document}
\newcommand{\TL}{\tilde{\mathcal{L}^{\mathcal{I}}}}
\newcommand{\LL}{\mathcal{L}^{\mathcal{I}}}
\newcommand{\KK}{\mathcal{K}_{\mathcal{I}}}
\newcommand{\iast}{\ast_{\mathcal{I}}}
\newcommand{\es}{\left( \begin{array}{c} g\ast\phi_s \\ \phi_s\end{array} \right)}
\newcommand{\ei}{\left(\begin{array}{c} g\ast\phi_i \\ \phi_i\end{array}
\right)}
\newcommand{\ees}{\left(\begin{array}{c} g\ast_{\mathcal{I}}\phi_s \\ 0 \end{array}
\right)}
\newcommand{\eei}{\left(\begin{array}{c} g\ast_{\mathcal{I}}\phi_i \\ 0 \end{array}
\right)}
\newcommand{\eees}{\left(\begin{array}{c} 0 \\ \phi_s\end{array}
\right)}
\newcommand{\eeei}{\left(\begin{array}{c} 0 \\ \phi_i\end{array}
\right)}
\topmargin 0pt
\headsep 0pt
\date{}
\title{Janossy Densities of Coupled Random Matrices}
\author{ 
Alexander Soshnikov\thanks{
Department of Mathematics,
University of California at Davis, 
One Shields Ave., Davis, CA 95616, USA.
Email address: soshniko@math.ucdavis.edu.
Research was supported in part by the Sloan Research Fellowship and the 
NSF grant DMS-0103948. 
}  }
\date{}
\maketitle
\begin{abstract}
We explicitly calculate the Janossy densities 
for a special class of
finite determinantal random point processes
with several types of particles
introduced by Pr\"ahofer and Spohn
and, in the full generality, by  Johansson in connection with  the analysis of polynuclear growth
models. The results of this paper generalize the theorem we proved earlier with Borodin
about the Janossy densities in biorthogonal ensembles.
In particular our results can be applied to ensembles of random matrices coupled in a chain which
provide a very important example of determinantal ensembles we study.
\end{abstract}

\section{Introduction and Formulation of Results}

{\it 1.1. Janossy densities.} The main goal of this paper is study the Janossy densities in a special class of finite determinantal random 
point processes introduced recently by 
by Pr\"ahofer and Spohn (\cite{PS}),
and, in the full generality, by Johansson (\cite{Jo1}) in connection with the analysis of a certain class of polynuclear growth models 
(the papers \cite{PS} and \cite{Jo1} considered sets of 
non-intersecting line ensembles with fixed initial 
and final points). A similar framework was introduced by Okounkov and Reshetikhin (\cite{OR}) and Ferrari and Spohn (\cite{FS}) to analyze 
the 3D Young tableaux via non-intersecting line ensembles. The distribution of the eigenvalues of random matrices coupled in a chain 
(see \cite{EM}, 
\cite{MMN}, \cite{E}, \cite{M}, \cite{AvM}, \cite{G})
also falls into this class.  The preprint \cite{FPS} shows that the the statistics of the random field 
built from the PNG model is related to the edge properties of a Gaussian multi-matrix model.

The term  Janossy densities in the theory of random point processes was introduced by Srinvasan in 1969 (\cite{Sr}) who referred to the 1950 
paper by Janossy
(\cite{Ja}) on  particle showers. The classical reference nowdays is 
(\cite{DVJ}, see chapters 5 and 7).
We postpone till section 1.2 the formal definition of the Janossy densities for the determinantal ensemble we are 
interested in. In order to give an uninitiated reader a glimpse of what is going on  let us consider
a random point process with one class of particles on $ \Bbb R \ $ (i.e. a probability measure on a space of 
locally finite point configuration in
$ \Bbb R $) and assume that all point correlation functions exist and locally integrable  (for all practical purposes the reader can think 
about a Poisson random point process for now). For a finite interval $I \subset \Bbb R \ $  we can define the k-point
Janossy density  $ \mathcal{J}_{k,I}(x_1, \ldots, x_k) \ $ for $ x_1, \ldots, x_k \in I \ $ as

\begin{eqnarray}
\label{Jan}
\mathcal{J}_{k,I}(x_1, \ldots,  
x_k) \* \prod_{i=1}^k dx_i = 
 & & \Pr \{ \textrm{
there are 
exactly $k$ particles  in  $I$
and  there is a particle in each }  \nonumber \\
& & \textrm{of the $k$  infinitesimal 
intervals} \ (x_i, x_i +dx_i), \  \ i=1, \ldots, k \}. 
\end{eqnarray}

A very useful property of the Janossy densities is that

\begin{equation}
\label{chislotochek}
\Pr \{ \textrm{there are exactly $  k $ particles in $I$} \} = 
\frac{1}{k!} \*
\int_{I^k}  \mathcal{J}_{k,I}
(x_1,\ldots, x_k)  \* dx_1
\cdots dx_k.
\end{equation}

In particular if we know that the number of the particles in a configuration is finite 
(which is often the case in Random Matrix Theory and other applications) one can use 
Janossy densities to study the  distribution of the largest/smallest particles (eigenvalues). Let us order the particles in a finite
configuration 
$ \lambda_1 \geq \lambda_2 \geq \lambda_3 \ldots \geq\lambda_n.$ 
Then the probability density of the $k-$th largest particle can be expressed in an integral form as
\begin{equation}
\label{janko}
\Pr(\lambda_{k} \in (s, s+ds))= \left( \frac{1}{(k-1)!} \*
\int_{(s, +\infty))^{k-1}}  \mathcal{J}_{k, (s, +\infty)}
(x_1,\ldots, x_{k-1}, s)  \* dx_1
\cdots dx_{k-1} \right)  \* ds,
\end{equation}
where in (\ref{janko}) we put $ x_k=s$.  The equivalent form of (\ref{janko}) is
\begin{equation}
\label{reno1}
\Pr(\lambda_{k} \geq s) - \Pr(\lambda_{k+1}\geq s) = \Pr (\#[s, +\infty) =k) = \frac{1}{k!} \* 
\int_{(s, +\infty)^{k}}  \mathcal{J}_{k, (s, +\infty)}
(x_1,\ldots, x_{k}) \* dx_1
\cdots dx_k.
\end{equation}

Janossy densities are equally useful if one is interested to study the local statistical properties of the distribution of particles
in the bulk of the configuration.

In (\cite{BoSo}) we studied  the Janossy densities in a so-called biorthogonal ensemble
(see e.g. \cite{Bor}  and 
\cite{TW}) which can be described as a random configuration of $n$ particles in a one-particle space $X$ (for simplicity we will consider 
the case $X=\Bbb R $) with the joint 
probability density (with respect to the Lebesgue measure in the case $X=\Bbb R $)
\begin{equation}
\label{bodensity}
 p_{n}(x_1, \ldots,  x_n) = \frac{1}{Z_{n}} \* \det(f_i(x_j))^n_{i,j=1} \*  \det(\phi_i(x_j))^n_{i,j=1},
\end{equation}
where $ \{f_i(x), \ \phi_i(x), \ i=1, \ldots, n \} \ $ are some complex-valued functions on $ \Bbb R.$
Such ensembles were extensively studied in random matrix theory (\cite{D}, \cite{M}), 
directed percolation and tiling models (\cite{Jo2}, \cite{Jo3}, \cite{Jo4}), models of 
uniform spanning trees and forests on graphs
(\cite{BP}, \cite{L})
and representation theory (\cite{BG}), among others.
For the biorthogonal ensemble one can calculate explicitely the $k-$ point correlation functions
\begin{equation}
\label{bocorr}
\rho_k^{(n)}(x_1,x_2, \ldots, x_k)= \frac{n!}{(n-k)!} \* \int_{\Bbb R^{n-k}} p_n(x_1, \ldots,  x_n) \* dx_{k+1}\* dx_{k+2} \cdots dx_n.
\end{equation}
It is a standrad fact that $k-$ point correlation functions have a determinantal form in this case
\begin{eqnarray}
\label{detform}
& & \rho_k^{(n)}(x_1,x_2, \ldots, x_k)=\det\big(K(x_i,x_j)\big)_{i,j=1, \ldots,k}, \\
\label{detform1}
& & K(x,y)= \sum_{i=1}^n \tilde{f_i}(x)\* \tilde{\phi_i}(y),
\end{eqnarray}
where $ \{\tilde{f_i}, \tilde{\phi_i}, i=1, \ldots,n \} $ are biorthogonal bases in $ Span\{f_i, \ 1\leq i \leq n\} $ and 
$ Span\{\phi_i, \ 1 \leq i \leq n\}, $ i.e.
\begin{equation}
\label{pairing}
\int  \tilde{f_i}(x) \* \tilde{\phi_j}(x) \* dx = \delta_{ij}.
\end{equation}

Then
the Janossy densities also have the
determinantal
form (see \cite{DVJ}, p.140 or \cite{BG}, Section 2) with a  kernel $ \mathcal{L}^I $ :
\begin{equation}
\label{yJanossy}
\mathcal{J}_{k,I}(x_1, \ldots, x_k)= 
const(I) \* 
\det(\mathcal{L}^I (x_i, x_j)_{i,j=1, \ldots,k}),
\end{equation}
where 
\begin{equation}
\label{Ll}
 \mathcal{L}^I= K_I (Id -K_I)^{-1}, \ \ \ const(I)= \det(Id-K_I),
\end{equation}
and the kernel of the integral operator $K_I$ is the restriction of the kernel of $K$ to $ I\times I, $ i.e.
$K_I(x,y)= \chi_I(x)\*K(x,y)\*\chi_I(y), \ $ where $ \chi_I $ is the indicator function of $I$. The probabilistic interpretation of
$cost(I)$ is that of the probability to have no particles in $I$.

The formula for the Janossy kernel $ \mathcal{L}^{I} $ for such 
ensembles was explicitely
calculated in (\cite{BoSo}). It was proved that the kernel $ \mathcal{L}^{I}$ can be constructed 
according to
the following rule:

{\it 1) consider the ensemble  with one-particle space $ X $ replaced by  $ X \setminus I $  (i.e.
the density of the distribution is still given by the same formula, only now it is defined on $(X \setminus I)^n,$
rather than  on $ X^n$; naturally the normalization constant changes).

2) calculate the correlation kernel on $ (X \setminus I)\times(X \setminus I)$ using (\ref{detform}),(\ref{detform1}) (i.e. the pairing
(\ref{pairing}) to be considered on the functions on $X \setminus I$).

3) extend the  correlation kernel to $ I \times I $ (since for $M=1 \ $ the correlation kernel is expressed in 
terms of
$ f_i, \ \phi_j \ $ there is no ambiguity in how to extend it to $ I \times I$).}

In the special case of a polynomial ensemble of Hermitian random matrices (which corresponds to
$ f_j(x)=\phi_j(x)=x^{j-1}\* \exp\left(-\frac{1}{2}\* V(x)\right), $ one can see that
the Janossy kernel can be written as the Christoffel-Darboux kernel
$ \mathcal{L}^{I}(x,y) = \sum_{j=0}^{n-1} p_j(x)\*p_j(y) $ built from the orthonormal polynomials with respect to the weight
$ \mu(dx)= \exp(-V(x))\* \chi_{I^c}(x) \* dx. $  
As one can see (\ref{janko}) and (\ref{reno1}) lead to a nice limiting expression for the distribution of the (appropriately rescaled) 
$k-$th largest
eigenvalue in the limit $ n \to \infty $ provided one can handle the asymptotics of the orthogonal polynomials with respect to the 
weight 
of the form $ e^{-n \* V(x)} \* \chi_{(-\infty, s)}(x) $ (where $s=s_n$ is at the  ``right edge of the spectrum''). Recent remarkable 
advances in
the application of Riemann-Hilbert problem technique to the asymptotics of orthogonal polynomials (see e.g. \cite{D}) suggest to us that
this might be possible to.
In particular in the GUE case the problem is thus reduced to the calculation of the asymptotics of the $n-$th orthonormal polynomial
with respect to the weight $ \exp(-x^2) \* \chi_{(-\infty, s_n)}(x), \ $ where $ s_n=2^{1/2}\* n^{1/2} + x \* 2^{-1/2}\* n^{-1/6} $ and
$ x $ is a fixed real number.  The limiting formulas 
suggested by (\ref{janko}) and (\ref{reno1}) would be,  in our opinion, simpler than the ones currently used that require $k$ 
differentiations of the Fredholm determinant of the infinite-dimensional integral operator  $ \det(1 + z \* K_{(s, +\infty)})$
with respect to the parameter $z$.
In a joint paper with Borodin (\cite{BoSo}) we were able to calculate explicitely the asymptotics of the $k-$th smallest eigenvalue in the 
standard 
Laguerre (Wishart) ensembles (see also \cite{Wie}).

In (\cite{S1}) we proved that the same recipe (with simple alterations) 
applies to pfaffian ensembles 
given by the formula (\cite{R}, \cite{O}, see also \cite{W})
\begin{equation}
\label{pfaff}
p(y_1, \ldots, y_{2n})=\frac{1}{Z_{2n}} \* \det(h_j(y_k))_{j,k=1, \ldots, 2n} 
\  pf (\epsilon(y_j, y_k))_{j,k=1, \ldots, 2n},
\end{equation}
where $ \ h_1, \ldots, h_{2n}$ are complex-valued functions on a measure space $ (Y, d\lambda(y)),$ and $ 
\epsilon(y,z)$ is a skew-symmetric kernel, $ \epsilon(y,z)=-\epsilon(z,y).$  For the definition of the pfaffian
of a $ 2n \times 2n $ skew-symmetric matrix we refer the reader to (\cite{Go}).  
The pfaffian ensemble (\ref{pfaff})
contains the biorthogonal ensemble as a specail case.  Some other interesting examples of the pfaffian ensembles include
$\beta=1$ (orthogonal) and $\beta =4 $ (symplectic) polynomial ensembles of random matrices (see \cite{S1}) as well as
the examples appearing in works on random growth models and vicious random walks (we refer to \cite{F} and \cite{NKT} and the 
references in there). The name pfaffian for the ensemble (\ref{pfaff}) comes from the fact that both  point correlation functions
and  Janossy densities have the pfaffian form (see \cite{R}, \cite{O}).

{\it 1.2. Determinantal ensembles with several classes of particles.}
Now we are ready to turn to the class of determinantal ensembles introduced in the papers by Johansson (\cite{Jo1})
and 
Pr\"ahofer and Spohn (\cite{PS}).
Let $ ( X , \mu) $ be a measure space, $ f_1, f_2, \ldots f_n, \ \phi_1, \phi_2, \ldots, \phi_n
 $ - complex-valued bounded integrable functions on 
$ X, $ and $ g_{1,2}(x,y), g_{2,3}(x,y), \ldots, g_{M-1,M}(x,y) $ -
complex-valued bounded integrable functions on $ X^2= X \times 
X $ with respect to the product measure $ \mu^{\otimes 2}=\mu \times \mu $ (in principle the above assumptions on
$ f_j, \ g_{l,l+1}, \phi_i, \ i,j=1, \ldots, n, \ l=1, \ldots, M $ can be weakened).  Suppose that
\begin{eqnarray}
\label{density}
& & p_{n,M}(x^{(1)}_1, \ldots, x^{(1)}_n; x^{(2)}_1, \ldots, x^{(2)}_n; 
\ldots; x^{(M)}_1, \ldots, x^{(M)}_n) \nonumber \\
&=& \frac{1}{Z_{n,M}} \* \det(f_i(x^{(1)}_j))_{i,j=1}^n \* \prod_{l=1}^{M-1}
\det(g_{l,l+1}(x^{(l)}_i, x^{(l+1)}_j))_{i,j=1}^n \* \det(\phi_j(x^{(M)}_i))_{i,j=1}^n
\end{eqnarray}
defines the density of a $ M \times n$- dimensional probability distribution
on $ X^{M \* n} = X \times \cdots \times X$ with respect to the 
product measure $ \mu^{\otimes M\* n}$. One can view the configuration \\ 
$ \overline{x}= 
(x^{(1)}_1, \ldots, x^{(1)}_n; x^{(2)}_1, \ldots, x^{(2)}_n; 
\ldots; x^{(M)}_1, \ldots, x^{(M)}_n) $ as the union of $ M $ configurations, namely
the first floor configuration $ \overline{x^{(1)}}=(x^{(1)}_1, \ldots, x^{(1)}_n)$,
the second floor configuration $ \overline{x^{(2)}}=(x^{(2)}_1, \ldots, x^{(2)}_n)$, etc. In other words we can
call the particles of the first floor  configuration   
- particles of the first class, the particles of the second floor configuration -  particles of the second 
class, etc. 
In the original papers (\cite{PS}), (\cite{Jo1}) the formulas of the type (\ref{density}) appeared as Pr\"ahofer,  Spohn
and Johansson considered sets of non-intersecting line ensembles with fixed initial and final points.

The normalization constant in (\ref{density})
(usually called the partition function) 
\begin{equation}
\label{partfun}
Z_{n,M}= 
\int_{X^{M\*n}} \* \det(f_i(x^{(1)}_j))_{i,j=1}^n \* \prod_{l=1}^{M-1}
\det(g_{l,l+1}(x^{(l)}_i, x^{(l+1)}_j))_{i,j=1}^n \* \det(\phi_i(x^{(M)}_j))_{i,j=1}^n
\* \prod_{l=1}^M \prod_{i=1}^n \mu(dx^{(l)}_i)
\end{equation}
can be shown to be equal $ (n!)^M \* \det(A), $ where the $ n \times n $ 
matrix $ A= (A_{jk})_{j,k=1, \ldots, n}$ is defined as
\begin{equation}
\label{A}
A_{jk}= \int_{X^M} f_j(x^{(1)}) \* \prod_{l=1}^{M-1} g_{l,l+1}(x^{(l)}, x^{(l+1)}) \phi_k(x^{(M)}) 
\prod_{m=1}^M \mu(dx^{(m)}).
\end{equation}
We assume that the matrix $ A $ is invertible.

It is easy to see that the biorthogonal ensemble (\ref{bodensity}) is a special case of (\ref{density}) corresponding to $ M=1.$
What is more interesting is that the ensemble 
(\ref{density}) in the case of two classes of particles (i.e. $M=2$) is a special case of the pfaffain 
ensemble (\ref{pfaff}) discussed at the end of section 1.1.   Indeed
let $Y$ be the disjoint union of two 
identical copies of $X, \ \ Y= X_1 \bigsqcup X_2, \ $ and the restriction of the measure $\lambda$ 
on each copy of $X$  given by $\mu$. Suppose that for $ 1\leq i \leq n $ the restriction of $ \  h_i \ $ 
on 
$ \ X_1 \ $ is given by $ \ f_i \ $ and the restriction of $ \  h_i \ $ on  $ X_2 \ $ is identically zero.
Similarly, suppose that the restriction of $ h_{n+i} $ on $ X_1$ is identically zero, and the restriction of 
$ h_{n+i} $ on $ X_2$ is given by $ \phi_i, \ \ i=1, \ldots, n.$  
Finally, suppose that the kernel $ \epsilon $ is identically zero
on $ X_1 \times X_1$ and $ X_2 \times X_2,$ and $ \epsilon(x_1, x_2)= -\epsilon(x_2, x_1) \ $ for 
$ \ x_1 \in X_1, \ \ x_2 \in X_2.$  Let us define a kernel $ g $ on $ X \times X $
which takes the same values as $ \epsilon$ on $ X_1 \times X_2. \ $
Then the formula (\ref{pfaff}) specializes into 
(\ref{density}),  $ \ \ M=2 \ $.

For the ensemble (\ref{density}) one can explicitly calculate 
$(k_1, k_2, \ldots, k_M)$-point correlation 
functions
\begin{equation}
\label{defcorr}
\rho_{k_1,\ldots,k_M}(x^{(1)}_1, \ldots, x^{(1)}_{k_1}; \ldots ; x^{(M)}_1, \ldots, x^{(M)}_{k_M}): = 
\int_{X^{M \* n -k}} \*
p_{n,M}(\bar{x}) \* \prod_{l=1}^M (n!/(n-k_l)!) \* \prod_{j=k_l+1}^n \* d\mu(x_j^{(l)}),
\end{equation}
where $ k=k_1+ \cdots + k_M,  0 \leq k_j \leq n, $ and show 
that they have the 
determinantal form (\cite{Jo1}, \cite{PS}, see also \cite{EM})

\begin{eqnarray}
\label{corr}
& & \rho_{k_1,\ldots,k_M}(x^{(1)}_1, \ldots, x^{(1)}_{k_1}; \ldots, x^{(M)}_1, \ldots, x^{(M)}_{k_M}) 
\nonumber \\ 
&=& \det(\mathcal{K}^{n,M}(l,x^{(l)}_{i_l}; m,x^{(m)}_{j_m}))_{l,m=1,\ldots,M, \ 1\leq i_l\leq k_l, 
1\leq i_m \leq k_m}.
\end{eqnarray}
To define the kernel $ \mathcal{K} $ we introduce the following notations for the convolutions:
\begin{eqnarray}
\label{convol}
& & g_{l,l+1} \ast g_{l+1,l+2}(x,y):= \int_X g_{l,l+1}(x,z) \* g_{l+1,l+2}(z,y) \* d\mu(z) \\
& & g_{l,m}:= g_{l,l+1} \ast \ldots \ast g_{m-1,m}, \ \ \ \ 1 \leq l < m \leq M \\
& & g_{l,m}:= 0, \ \ \ 1 \leq m \leq l \leq M.
\end{eqnarray}
We will use similar notations for the integrals $ \int_X f_j(x) \* g_{1,m}(x,y) \* d\mu(y) $ and 
$ \int_X g_{m,M}(x,y) \* \phi_s(y) \* d\mu(y), $ namely
\begin{eqnarray}
\label{fgconvol}
& & (f_j \ast g_{1,m})(y)= \int_X f_j(x) \* g_{1,m}(x,y) \* d\mu(x), \ \ \ j=1, \ldots,n,  \\
& & (g_{m,M} \ast \phi_s)(x)=\int_X g_{m,M}(x,y) \* \phi_s(y) \* d\mu(y), \ \ \ s=1, \ldots, n.
\end{eqnarray}
The kernel $ \mathcal{K}^{n,M} : ( \{1,2,\ldots,M \})\times X)^2 \mapsto \Bbb C $ is a $ M \times M $ matrix kernel 
given by the following expression 
\begin{equation}
\label{K}
\mathcal{K}^{n,M}(l,x; m,y)=  - g_{lm}(x,y) + \sum_{i,j=1}^n (g_{l,M} \ast \phi_i)(x) \* (A^{-1})_{ij} \* 
(f_j \ast g_{1,m})
\end{equation}
(to simplify the  formulas we adopted the convention $ \ f_i \ast g_{1,m} = f_i \ $ for $ \ m=1 $ and
$ \ g_{l,M} \ast \phi_j = \phi_j \ $ for $ \ l=M ). $
Usually we  omit the dependence on $ n $ and $ M $  in the notation of the kernel  if it does not lead to
ambiguity. 

{\bf Remark}

{\it Repeated applications of the Heine identity
$$ \frac{1}{n!} \* \int_{X^n} \det(\varphi_i(x_j))_{i,j=1, \ldots, n} \* \det(\psi_i(x_j))_{i,j=1,\ldots,n} \* 
\prod_{k=1}^n d\mu(x_k) = \det \bigl( \int_X \varphi_i(y) \* \psi_j(y) \* d\mu(y) \bigr)_{i,j=1}^n $$
to (\ref{density}) implies that the joint distribution of the $ l_1, \ldots, l_m$- floors configurations 
$ \overline{x^{(l_1)}}, \ldots, \overline{x^{(l_m)}},
\ \ 1 \leq l_1 < \ldots l_m \leq M, $  is again of determinantal form (\ref{density})
with $ \tilde{f_i}=f_i\ast g_{1, l_1}, \ \tilde{g}_{1,2}= g_{l_1, l_2}, \ \tilde{g}_{2,3}= g_{l_2,l_3},
\ldots, \tilde{g}_{m-1, m}= g_{l_{m-1}, l_m}, \ \tilde{\phi_j}= g_{l_m,M}\ast \phi_j, \ \ \tilde{M}=m \ $  and
$ \ \tilde{A}=A.$}

If $ X \subset \Bbb R $ and $\mu $ is absolutely continuous with respect 
to the Lebesgue measure, then the probabilistic meaning of the $(k_1, \ldots, k_M)$ -point 
correlation functions is that of the 
density of probability to find a particle of the first class in each infinitesimal
interval around points $ x^{(1)}_1, \ldots, x^{(1)}_{k_1},$  a particle of the second class
in each infinitesimal interval around points $ x^{(2)}_1, \ldots, x^{(2)}_{k_2}, 
$ etc. In other words
\begin{eqnarray*}
& & \rho_{k_1,\ldots, k_M}(x^{(1)}_1, \ldots, x^{(1)}_{k_1}; \ldots; x^{(M)}_1, \ldots, 
x^{(M)}_{k_M}) \mu(dx^{(1)}_1) \cdots 
\mu(dx^{(M)}_{k_M})=\\ 
& &\Pr  \bigl\{ \textrm{ for each $ 1 \leq l \leq M $ there is a particle of the $l$-th class in each  
of the $k_l$ intervals} \\ 
& & (x^{(l)}_i, x^{(l)}_i +dx^{(l)}_i),  \ 1 \leq i \leq k_l \bigr\}. 
\end{eqnarray*}

On the other hand, if $\mu$ is supported by a discrete set of points, then
\begin{eqnarray*}
& & \rho_{k_1,\ldots, k_M}(x^{(1)}_1, \ldots, x^{(1)}_{k_1}; \ldots; x^{(M)}_1, \ldots, x^{(M)}_{k_M}) 
\mu(x^{(1)}_1) \cdots 
\mu(x^{(M)}_{k_M})= \\
& & \Pr \, \bigl\{ \textrm{ for each $ 1 \leq l \leq M $ there is a particle of the $ l$-th class at each of the sites } \ \ 
x^{(l)}_i,  \ i=1, \ldots, k_l \bigr\}. 
\end{eqnarray*}

In general, random point processes with the point correlation functions
of the determinantal form are called determinantal (a.k.a. fermion)
random point processes
(\cite{S2}).

The Janossy densities $ \mathcal{J}_{k_1,I_1; k_2, I_2; \ldots, k_M, I_M}(x^{(1)}_1, \ldots,  
x^{(1)}_{k_1}; \ldots; x^{(M)}_1, \ldots, x^{(M)}_{k_M}),
\ \ 0 \leq k_l \leq n, \  l=1, \ldots, M, \ \ k_1 + \ldots + k_M = k \leq M \times n, \ $
describe the joint distribution of the first class particles  in  $I_1$, second class particles in $I_2$, ..., 
$ M$-th class particles in $I_k$, 
where $ I_1, I_2, \ldots, I_k $ are measurable subsets of 
$ X. \ $  
For the ensembles with a finite number of particles Janossy densities can be obtained from the joint 
probability distribution of  the particles by integration, in particular for (\ref{density}) we have

\begin{eqnarray}
\label{defJanossy}
& & \mathcal{J}_{k_1,I_1; k_2, I_2; \ldots, k_M, I_M}(x^{(1)}_1, \ldots,  
x^{(1)}_{k_1}; \ldots; x^{(M)}_1, \ldots, x^{(M)}_{k_M}) := \nonumber \\
& & \int_{(X\setminus I_1)^{k_1} \times \cdots \times (X \setminus I_M)^{k_M}} \*
p_{n,M}(\bar{x}) \* \prod_{l=1}^M (n!/(n-k_l)!) \* \prod_{j=k_l+1}^n \* d\mu(x_j^{(l)}),
\end{eqnarray}
One can say that the Janossy density $ \mathcal{J}_{k_1,I_1; k_2, I_2; \ldots; k_M, I_M}(x^{(1)}_1, \ldots,  
x^{(1)}_{k_1}; \ldots; x^{(M)}_1, \ldots, x^{(M)}_{k_M})$
gives the joint density of the distribution of  $ k_1 $ first class particles  
in $ I_1$,  $k_2$ second class particles in $I_2$, ...,  $k_M \ $ $ \ M$-th class particles
in $I_M$ (under the assumption that there are no other particles of the first class in $ I_1$, 
no other particles 
of the second kind in $ I_2$, etc). The Janossy densities differ from the the conditional 
probability densities by the normalization: the Janossy  densities are normalized in such a way that the whole 
mass 
is not one, but rather
\begin{eqnarray*}
\label{intjan}
& & 
\frac{1}{k_1 ! \cdots k_M !}  \*
\int_{I^{k_1}_1\times \ldots \times I^{k_M}_M} \*  \mathcal{J}_{k_1,I_1;\ldots; k_M, I_M}(x^{(1)}_1, 
\ldots, x^{(1)}_{k_1}; 
\ldots, x^{(M)}_1, \ldots, x^{(M)}_{k_M}) \* \prod_{l=1}^M \* \prod_{i_l=1}^{k_M} d\mu(x^{(l)}_{i_l}) = \\
& & \Pr \,  \{\textrm{
for each $ 1 \leq l \leq M $ there are 
exactly $k_l$
particles of the $l-$th class in  $I_l$} \}
\end{eqnarray*}

Let $ x^{(1)}_1, \ldots, x^{(1)}_{k_1} $ be some distinct 
points of $ I_1, \ \  x^{(2)}_1, \ldots, x^{(2)}_{k_2} \  $ 
- some distinct points of $ I_2, \ \ \ldots,  x^{(M)}_1, \ldots, x^{(M)}_{k_M} \ $ - some distinct points of 
$I_M.$
If
$ X \subset \Bbb R $ and $ \mu $ is absolutely continuous with respect to
the Lebesgue measure, then  
\begin{eqnarray*}
& & \mathcal{J}_{k_1,I_1; k_2, I_2; \ldots, k_M, I_M}(x^{(1)}_1, \ldots, 
x^{(1)}_{k_1}; \ldots; x^{(M)}_1, \ldots, x^{(M)}_{k_M})
\* \mu(dx^{(1)}_1) \cdots \mu(dx^{(M)}_{k_M}) =\\
& &  \Pr \,  \{\textrm{
there are 
exactly $k_1$
particles of the first class in  $I_1$, \ldots, $k_M$ particles of the $ M$-th class in $ I_M$,} \\
& & \textrm{so that  for each  $ 1 \leq l \leq M $  there is a particle of the $l$-th class in each 
of the $k_l$  infinitesimal} \\ 
& & \textrm{intervals}(x^{(l)}_{i}, x^{(l)}_{i} +dx^{(l)}_{i}), \  \ i=1, \ldots, k_l \}. 
\end{eqnarray*}
Similarly, if  $ \mu $ is discrete, then
\begin{eqnarray*}
& & \mathcal{J}_{k_1,I_1; k_2, I_2; \ldots, k_M, I_M}(x^{(1)}_1, \ldots, 
x^{(1)}_{k_1}; \ldots; x^{(M)}_1, \ldots, x^{(M)}_{k_M})
\* \mu(x^{(1)}_1) \cdots \mu(x^{(M)}_{k_M}) =\\
& & \Pr \,  \{\textrm{
there are 
exactly $k_1$
particles of the first class in  $I_1$,  \ldots, $k_M$ particles of the $M$-th class in $ I_M$,} \\
& & \textrm{ so that for each $ 1 \leq l \leq M $ there is a particle of the $l$-th class  at each
of the $k_l$ sites}\\
& &  x^{(l)}_{i},  \ \  \ i=1, \ldots, k_l \}.
\end{eqnarray*}
See \cite{DVJ}, \cite{BoSo} and \cite{S2} for additional discussion. It is instructive to compare the 
probabilistic interpretation of the Janossy densities with the probabilistic interpretation of the correlation 
functions given above.
For determinantal processes
the Janossy densities also have the
determinantal
form (see \cite{DVJ}, p.140 or \cite{BG}, Section 2) with a  kernel $ \mathcal{L}^{\mathcal{I}}$ :
\begin{eqnarray}
\label{Janossy}
& & \mathcal{J}_{k_1,I_1;\ldots; k_M, I_m}(x^{(1)}_1, \ldots, x^{(1)}_{k_1}; 
\ldots, x^{(M)}_1, \ldots, x^{(M)}_{k_M}) \nonumber \\ 
&=& const(\mathcal{I}) \* 
\det(\mathcal{L}^{\mathcal{I}} (l,x^{(l)}_{i_l}; m,x^{(m)}_{j_m}))_{l,m=1,\ldots,M, \ 1\leq i_l\leq k_l, \ 
1\leq i_m \leq k_m},
\end{eqnarray}
where 
\begin{equation}
\label{L}
 \mathcal{L}^{\mathcal{I}}= \mathcal{K}_{\mathcal{I}}(Id -\mathcal{K}_{\mathcal{I}})^{-1},
\end{equation}
and the notations $\mathcal{I},   \ const(\mathcal{I})$ and 
$ \mathcal{K}_{\mathcal{I}} $ are explained in the next paragraph. One can point out that we already have seen formulas of this type when we 
discussed a special case of biorthogonal ensembles.

 The integral operator $\mathcal{K}$ 
acts on a Hilbert space $\mathcal{H},$  which is the orthogonal direct sum of 
$ M $ copies of $ L^2(X, \mu), $ i.e. $ \mathcal{H}=L^2(X, \mu) \bigoplus \ldots \bigoplus L^2(X,\mu). \  
$ Let $ \mathcal{X} $ be the disjoint union of $ M $ identical copies of $ X $, in other words 
$ \mathcal{X}= X_1 \bigsqcup \ldots \bigsqcup X_M,$ where each $X_l, \ l=1, \ldots, M, $ is a copy of $ X. \ $
One can think of
$ X_l, \ 1\leq l \leq M, \ $ being the $l$-th floor in our particle space. 
Extending the measure $ \mu $ in a natural way to $ \mathcal{X}$ and denoting the extension by $ \mu_M $
we can view $ \mathcal{H} $ as the Hilbert space
$ L^2(\mathcal{X}, \mu_M).$  For $ I_1 \subset X, \ldots, I_M \subset X,$  we construct a 
subset of the particle space $ \mathcal{X}$ , 
denoted by $ \mathcal{I} $, in such a way that the 
intersection of $ \mathcal{I} $ with $X_l$  is equal to $ I_l, \  l=1, \ldots, M$. Let us denote
by $ \mathcal{K}_{\mathcal{I}} $ the restriction of the integral operator $ \mathcal{K} $ 
to $ L^2(\mathcal{I}, \mu_M) = L^2(I_1) \bigoplus \ldots \bigoplus L^2(I_M)$ 
(in other words we restrict the kernel $ \mathcal{K} $ 
to $ \mathcal{I} \times \mathcal{I}.$) The normalization constant
$const(I_1, \ldots, I_M)= const(\mathcal{I})$ is given by the Fredholm determinant  
$ const(\mathcal{I})= \det(Id - \mathcal{K}_{\mathcal{I}})$ 
of the operator $ \mathcal{K}_{\mathcal{I}}$  
(for the definition of the Fredholm determinant we refer the reader to 
\cite{RS}, \cite{Si}).  The probabilistic meaning of the normalization constant
$ const(\mathcal{I})$ is that of the probability  to have no first class particles in $ I_1,$ 
no second class particles in $ I_2,$ ...., no $M$-th class particles in $I_M.$

{\bf Remark}

{\it  Strictly speaking,  to view the ensemble (\ref{density}) as a determinantal random point process
we have to consider it as a distribution of $ \ M \times n \ $  identical particles on a one-particle space
$\mathcal{X} \ $, so that with probability $ 1 $  there are exactly $n$ particles in each 
$ \ X_l, \ \ l=1, \ldots, M, $ and the joint distribution of the particles in $X_1, \ X_2, \ldots, X_M $ 
is given by (\ref{density}). Then the $(k_1, k_2, \ldots, k_M) $ -point correlation function is nothings else 
but a usual $k$-point correlation function, $ k=k_1+\cdots +k_M, \ $ and $k_l \  $ 
is just the number of arguments of the $k$-point correlation function that belong to $X_l, \ l=1, \ldots, M $.
In particular one can easily see that
\begin{equation}
\label{almostproj}
\sum_{l=1}^M \int_X \mathcal{K}(k,x;l,y) \* \mathcal{K}(l,y;m,z) \* d\mu(y) = \mathcal{K}(k,x; m,z) + (m -k) \* \mathcal{K}(k,x; m,z)
+ 2(m-k) \* g_{k,m}(x,z)
\end{equation}
and an easy generalization of the Dyson-Mehta lemma (\cite{M},  Theorem 5.2.1) claims that if we integrate out
(over $\mathcal{X} $ ) a variable in the determinant of the $ k\times k$ matrix with the correlation kernel
$\mathcal{K}$ we obtain (up to a trivial combinatorial coefficient) the determinant of the 
$(k-1)\times (k-1)$ matrix with the same kernel. To the reader familiar with the original Dyson-Mehta argument 
we note that the terms with the factor $(m-k)$  vanish since for any 
permutation $\sigma \in S_k \ $ one (trivially) has $ \sum_{l=1}^k (\sigma(l) - \sigma(l+1)) =0, \ $ where we put
$ \sigma(k+1):= \sigma(1). \ $
}

{\it 1.3. Formulation of the main result.}
If one can prove that the analogue
the recipe for the Janossy kernel described above in the biorthogonal case (after formulas ((\ref{yJanossy}), (\ref{Ll}))  applies to
the ensemble (\ref{density}) for general $ M, \  $ the Janossy kernel would have the following form:
\begin{equation}
\label{Lform}
\mathcal{L}^{\mathcal{I}}(l,x; m,y)=  - g^c_{lm}(x,y) + 
\sum_{i,j=1}^n (g^{c}_{l,M} \ast_c \phi_i)(x) \* (A^{c})^{-1}_{ij} \* 
(f_j \ast_c g^c_{1,m})(y),
\end{equation}
where
\begin{eqnarray}
\label{Icconvol1}
& & g_{l,l+1} \ast_{c} g_{l+1,l+2}(x,y):= \int_{I^c_{l+1}} 
g_{l,l+1}(x,z) \* g_{l+1,l+2}(z,y) \* d\mu(z) \\
\label{Iconvol2}
& & g^{c}_{l,m}:= g_{l,l+1} \ast_{c} \ldots \ast_{c} g_{m-1,m}, 
\ \ \ \ 1 \leq l < m \leq M \\
& & g^{c}_{l,m}:= 0, \ \ \ 1 \leq m \leq l \leq M, 
\end{eqnarray}
\begin{eqnarray}
\label{Iconvol3}
& & (f_j \ast_{c} g^c_{1,m})(y)= \int_{I^c_1} f_j(x) \* g^c_{1,m}(x,y) \* d\mu(x), \ \ \ j=1, \ldots,n,  \\
\label{Iconvol4}
& & (g^c_{m,M} \ast_{c} \phi_s)(x)=\int_{I^c_M} g^c_{m,M}(x,y) \* \phi_s(y) \* d\mu(y), \ \ \ s=1, \ldots, n, \\
\label{Iconvol5}
& & A^c_{jk}= \int_{I^c_1 \times \ldots \times I^c_M} 
f_j(x^{(1)}) \* \prod_{l=1}^{M-1} g_{l,l+1}(x^{(l)}, x^{(l+1)}) \phi_k(x^{(M)}) 
\prod_{m=1}^M \mu(dx^{(m)}),
\end{eqnarray}
(the notation $ I^c$ stands for the complement of a set $ I$, we also remind the reader that we use the 
convention $ f_j \ast_c g^c_{1,m} = f_j \ $ for $ m=1 \ $ and $ \ g^c_{l,M}\ast_c \phi_s = \phi_s \ $ for 
$ \ l=M. $)

Throughout the paper we  assume that the matrices  $ A$ (defined in (\ref{A})), 
$ A^c \ $ (defined in (\ref{Iconvol5})) and $ A^{\mathcal{I}} $ (defined below in (\ref{igogo}))
are invertible.
The main result of this paper is

\begin{theorem}
 The Janossy kernel $\mathcal{L}^{\mathcal{I}}$  in the determinantal ensemble (\ref{density}) is given by the formulas 
(\ref{Lform}-\ref{Iconvol5}) for all $M$.
\end{theorem}

To finish this section we introduce the notations for the convolutions over $ I_1, \ldots, I_M.$
\begin{eqnarray}
\label{Icconvol}
& & g_{l,l+1} \ast_{\mathcal{I}} g_{l+1,l+2}(x,y):= \int_{I_{l+1}} 
g_{l,l+1}(x,z) \* g_{l+1,l+2}(z,y) \* d\mu(z) \\
& & (f_j \ast_{\mathcal{I}} g_{1,m})(y)= \int_{I_1} f_j(x) \* g_{1,m}(x,y) \* d\mu(y), \ \ \ j=1, \ldots,n,  \\
& & (g_{m,M} \ast_{\mathcal{I}} \phi_s)(x)=\int_{I_M} g_{m,M}(x,y) \* \phi_s(y) \* d\mu(y), 
\ \ \ s=1, \ldots, n, \\
\label{igogo}
& & A^{\mathcal{I}}_{jk}= \int_{I_1 \times \ldots \times I_M} 
f_j(x^{(1)}) \* \prod_{l=1}^{M-1} g_{l,l+1}(x^{(l)}, x^{(l+1)}) \phi_k(x^{(M)}) 
\prod_{m=1}^M \mu(dx^{(m)}).
\end{eqnarray}

The case $ M=1$ of the Theorem was proven in 
\cite{BoSo}. The case $ M=2$ (``two-matrix model'') follows from our proof  for the pfaffian 
ensembles (\ref{pfaff}) given in (\cite{S1}). 
The case $ M=3$
(``three-matrix model'')  will be proven in section 3 as a ``warm up''. The proof of the Theorm for general $M$ 
is given in section 4.
We devote section 2 to some additional examples of the determinantal ensembles (\ref{density}).

\section{Examples of Determinantal Ensembles}

{\bf One Matrix Models. Unitary Ensembles}

Let $ M=1$. In the special cases $X=\Bbb R, \ f_j(x)=\phi_j(x)=x^{j-1}, $ and $ X=
\{\Bbb C\,|\, |z|=1 \}, f_j(z) =\overline{\phi}_j(z)=
z^{j-1}$, such ensembles are well known in Random Matrix Theory
as {\it unitary ensembles}, see \cite{M} for details. An ensemble  of the 
form (\ref{bodensity}) which is different from random matrix ensembles was
studied in \cite{Mut}.  

%
%

{\bf Matrices Coupled in a Chain}

Consider the chain of $M$ complex Hermitian $ n\times n$ matrices with the joint probability density
(with respect to 
the $M \times n^2$-dimensional Lebesgue measure $ \prod_{l=1}^M dA_l$) given by the formula (see \cite{M}, 
\cite{EM})
\begin{eqnarray}
\label{multimatrix}
& & F(A_1,\ldots, A_M)= const(n,M) \* 
\exp\left( -Tr\bigl\{ \frac{1}{2} V_1(A_1)+ V_2(A_2)+\cdots +V_{M-1}(A_{M-1})+
\frac{1}{2}V_M(A_M)\bigr\}\right) \nonumber \\
& & \times \exp \left( Tr \{c_1 \*A_1 \* A_2+ c_2\* A_2\* A_3+\cdots +c_{M-1}\* A_{M-1}\* A_M \}\right).
\end{eqnarray}
The case $ M=1$ corresponds to the one matrix model discussed above.
Let us denote by $ \lambda^{(l)}_1, \lambda^{(l)}_2, \ldots, \lambda^{(l)}_n $ the eigenvalues 
(all real) of $ A_l, 
\ l=1, \ldots,n.$ The probability density of the joint distribution of the eigenvalues of $ A_1, \ldots, A_M$
with respect to the Lebesgue measure on $\Bbb R^{M\*n}$ is given by the formula (\cite{M}, \cite{EM})
\begin{eqnarray}
\label{multeigen}
& & p(x^{(1)}_1, \ldots, x^{(M)}_n)=\frac{1}{Z_{n,M}} \* \exp 
\left( -\sum_{i=1}^n \bigl\{ \frac{1}{2} \* V_1(x^{(1)}_i)
+ V_2(x^{(2)}_i) + \ldots + V_{M-1}(x^{(M-1)}_i)+ \frac{1}{2}\* V_M(x^{(M)}_i) \bigr\} \right) \nonumber \\
& \times & \prod_{1\leq i < j \leq n} \* (x^{(1)}_i - x^{(1)}_j) \* (x^{(M)}_i - x^{(M)}_j) \* 
\prod_{l=1}^{M-1}
\det(\exp(c_l\*x^{(l)}_i x^{(l+1)}_j))_{i,j=1}^n .
\end{eqnarray}
Writing the  Vandermonde products in (\ref{multeigen}) as determinants we arrive at the expression of the form
(\ref{density}). In should be noted that while in the case of one-matrix polynomial ensembles the existing Riemann-Hilbert problem machinery
(see e.g. \cite{D}) could,  in principle, provide the desired asymptotics of the orthogonal polynomials with respect to the weights
$ \exp(-n\* V(x)) \* \chi_{I_n^c}(x), $ this is not yet the case  for multi-matrix models.

{\bf Non-Intersecting Paths of a Markov Process}

We follow \cite{KMcG}, \cite{Jo5}. Let $ p_{t,s}(x,y)$ be the transition probability of a Markov process
$\xi(t)$ on $ \Bbb R $ with continuous trajectories and  $ (\xi_1(t), \xi_2(t), \ldots, \xi_n(t))$ - 
$ n $  independent copies of the process. A beautiful classical result of Karlin and McGregor states that if  
$n $ particles start at the positions $ x^{(0)}_1< x^{(0)}_2< \ldots< x^{(0)}_n, $ then the probability density 
of their joint distribution at time $ t_1 >0, $ given that their paths have not intersected
for all $ 0\leq t \leq t_1, $ is  equal to 
$$ \pi_{t_1}(x^{(1)}_1, \ldots, x^{(1)}_n)=\det(p_{0,t_1}(x^{(0)}_i, x^{(1)}_j))_{i,j=1}^n $$
provided the process $ (\xi_1(t), \xi_2(t), \ldots, \xi_n(t))$ in $ \Bbb R^n $ has a strong Markovian 
property. To understand the above formula better one can consider first the case of two particles and use a 
standard
reflection trick to check that the result is correct. The most general combinatorial form of Karlin-McGregor theorem is known as the 
Gessel-Viennot theorem (\cite{GV}), we refer the reader for the additional discussion to \cite{Stan}), section 2.7.
 
Let $ 0<t_1<t_2< \ldots <t_{M+1}.$ The conditional probability density that the particles are in the 
positions $  x^{(1)}_1< x^{(1)}_2< \ldots< x^{(1)}_n $ at time $ t_1 $, at the positions
$  x^{(2)}_1< x^{(2)}_2< \ldots< x^{(2)}_n $ at time $ t_2 $,..., at the positions
$  x^{(M)}_1< x^{(M)}_2< \ldots< x^{(M)}_n $ at time $ t_M $, given that at time $ t_{M+1}$ 
they are at the positions $  x^{(M+1)}_1< x^{(M+1)}_2< \ldots< x^{(M+1)}_n $  and their paths have not 
intersected, is then equal to
\begin{equation}
\label{nonintersect}
 \pi_{t_1, t_2, \ldots, t_M}(x^{(1)}_1, \ldots, x^{(M)}_n)= \frac{1}{Z_n,M} \* \prod_{l=0}^M \*
\det(p_{t_l,t_{l+1}}(x^{(l)}_i, x^{(l+1)}_j))_{i,j=1}^n , 
\end{equation}
where $ t_0 =0.$ One can easily see that (\ref{nonintersect}) belongs to the
class of ensembles
$(\ref{density})$.
As an interesting related example we refer to the random walks on a discrete circle (see \cite{F} and 
\cite{Jo1}, section 2.3)

Finally we refer to (\cite{Jo1} and \cite{PS}, specifically to the formulas  (1.17)-(1.19), (3.15)-(3.16) in 
the first reference) for an example of a determinantal ensemble
(\ref{density}) appearing in the analysis of a  polynuclear growth model.

\section{Case of Three Classes of Particles.}

We devote this section to the proof of our main result in the special case of three classes of particles.
In the case $ M=3$ the formulas (\ref{density}- \ref{A}), (\ref{K}) have the following form:

\renewcommand{\es}{\left( \begin{array}{c} g_{12}\ast g_{23} \ast \phi_s \\ g_{23} \ast 
\phi_s \\ \phi_s \end{array} \right)}
\renewcommand{\ei}{\left( \begin{array}{c} g_{12}\ast g_{23} \ast \phi_i \\ g_{23} \ast 
\phi_i \\ \phi_i \end{array} \right)}
\renewcommand{\ees}{\left(\begin{array}{c} g_{12}\ast_{\mathcal{I}} g_{23} \ast \phi_s \\ g_{23} 
\ast_{\mathcal{I}} \phi_s \\ 0 \end{array}
\right)}
\renewcommand{\eei}{\left(\begin{array}{c} g_{12}\ast_{\mathcal{I}} g_{23} \ast \phi_i \\ g_{23} 
\ast_{\mathcal{I}} \phi_i \\ 0 \end{array}
\right)}
\renewcommand{\eees}{\left(\begin{array}{c}  g_{12} \ast g_{23} \ast_{\mathcal{I}} \phi_s \\ 0 \\ 0 \end{array}
\right)}
\renewcommand{\eeei}{\left(\begin{array}{c}  g_{12} \ast g_{23} \ast_{\mathcal{I}} \phi_i \\ 0 \\ 0 \end{array}
\right)}
\newcommand{\eeees}{\left(\begin{array}{c} g_{12} \ast_{\mathcal{I}} g_{23} \ast_{\mathcal{I}} \phi_s \\ 0 
\\ 0 \end{array}
\right)}
\newcommand{\eeeei}{\left(\begin{array}{c} g_{12} \ast_{\mathcal{I}} g_{23} \ast_{\mathcal{I}} \phi_i \\ 0 
\\ 0 \end{array}
\right)}
\newcommand{\Fs}{\left(\begin{array}{c} 0 \\ g_{23} \ast \phi_s \\ \phi_s
\end{array}
\right)}
\newcommand{\Fi}{\left(\begin{array}{c} 0 \\ g_{23} \ast \phi_i \\ \phi_i
\end{array}
\right)}
\newcommand{\FFFs}{\left(\begin{array}{c} 0 \\ g_{23} \ast \phi_s \\ 0
\end{array}
\right)}
\newcommand{\FFFi}{\left(\begin{array}{c} 0 \\ g_{23} \ast \phi_i \\ 0
\end{array}
\right)}
\newcommand{\FFs}{\left(\begin{array}{c} 0 \\ g_{23} \ast_{\mathcal{I}} \phi_s \\ 0
\end{array}
\right)}
\newcommand{\FFi}{\left(\begin{array}{c} 0 \\ g_{23} \ast_{\mathcal{I}} \phi_i \\ 0
\end{array}
\right)}

\begin{eqnarray}
\label{density3}
& & p_{n,3}(x^{(1)}_1, \ldots, x^{(1)}_n; x^{(2)}_1, \ldots, x^{(2)}_n; 
x^{(3)}_1, \ldots, x^{(3)}_n) \nonumber \\
&=& \frac{1}{Z_{n,3}} \* \det(f_i(x^{(1)}_j))_{i,j=1}^n \* 
\det(g_{1,2}(x^{(1)}_i, x^{(2)}_j))_{i,j=1}^n \* 
\det(g_{2,3}(x^{(2)}_i, x^{(3)}_j))_{i,j=1}^n  \*
\det(\phi_j(x^{(3)}_i))_{i,j=1}^n\\
\label{partfun2}
& & Z_{n,3}= \det A, \\
\label{A3}
& & A_{ij}= \int_{X \times X \times X } f_i(x) \* g_{1,2}(x,y) \* g_{2,3}(y,z) \phi_j(z) 
\* d\mu(x) \* d\mu(y) \* d\mu(z)= f_i \ast g_{1,2} \ast g_{2,3} \ast \phi_j.
\end{eqnarray}
Let us write $ \mathcal{K} $ and $ \TL $  as  $ 3 \times 3 $ matrix kernels
\begin{eqnarray}
\label{K3}
& &  \mathcal{K}(x,y)= \\
& & \sum_{i,j=1}^n A^{-1}_{ij} 
\left( \begin{array}{ccc}
(g_{1,2}\ast g_{2,3} \ast \phi_i)\otimes f_j & (g_{1,2} \ast g_{2,3} \ast\phi_i) \otimes
(f_j \ast g_{1,2})  & (g_{1,2} \ast g_{2,3} \ast \phi_i)\otimes (f_j \ast g_{1,2} \ast g_{2,3}) \\
(g_{2,3} \ast \phi_i)\otimes f_j & (g_{2,3} \ast\phi_i) \otimes
(f_j \ast g_{1,2})  & (g_{2,3} \ast \phi_i)\otimes (f_j \ast g_{1,2} \ast g_{2,3}) \\
\phi_i\otimes f_j & \phi_i \otimes
(f_j \ast g_{1,2})  & \phi_i\otimes (f_j \ast g_{1,2} \ast g_{2,3}) 
\end{array} \right) \nonumber \\
& + & \left(\begin{array}{ccc} 0 & g_{1,2} & g_{1,2} \ast g_{2,3} \\ 0 & 0 & g_{2,3}  \\
0 & 0 & 0 
\end{array} \right), \nonumber 
\end{eqnarray}
\begin{eqnarray}
\label{L3}
& &  \TL(x,y)= 
\sum_{i,j=1}^n (A^c)^{-1}_{ij} \times \\
& & \left( \begin{array}{ccc}
(g_{1,2}\ast_c g_{2,3} \ast_c \phi_i)\otimes f_j & (g_{1,2} \ast_c g_{2,3} \ast_c \phi_i) \otimes
(f_j \ast_c g_{1,2})  & (g_{1,2} \ast_c g_{2,3} \ast_c \phi_i)\otimes (f_j \ast_c g_{1,2} \ast_c g_{2,3}) \\
(g_{2,3} \ast_c \phi_i)\otimes f_j & (g_{2,3} \ast_c \phi_i) \otimes
(f_j \ast_c g_{1,2})  & (g_{2,3} \ast_c \phi_i)\otimes (f_j \ast_c g_{1,2} \ast_c g_{2,3}) \\
\phi_i\otimes f_j & \phi_i \otimes
(f_j \ast_c g_{1,2})  & \phi_i\otimes (f_j \ast_c g_{1,2} \ast_c g_{2,3}) 
\end{array} \right) \nonumber \\
&+& \left(\begin{array}{ccc} 0 & g_{1,2} & g_{1,2} \ast_c g_{2,3} \\ 0 & 0 & g_{2,3} \\
0 & 0 & 0 
\end{array} \right). \nonumber
\end{eqnarray}
Our goal is to show that $ \TL $ is equal to $ \LL= \KK \* (Id - \KK)^{-1} \ $ on $ L^2(\mathcal{I}).$ The main 
idea is to use the fact that
(for arbitrary $ M $ ) $ \KK$ is ``almost'' a finite rank operator. Namely, it is equal to a sum of a finite 
rank operator and a nilpotent operator. We first check the identity $ \TL= \LL$ on a sufficiently large 
finite-dimensional subspace. 
Let us introduce  a finite-dimensional subspace  of $ L^2(\mathcal{I}),$
$$ W= Span \left\{  \ei, \ \eei, \ \eeei, \ \eeeei \right \}_{i=1, \ldots, n}.$$
We claim that  both $\KK$ and $\TL$ leave $ W $ invariant.

\begin{lemma}
The operators $ \KK, \ \ \TL$ leave
$W$ invariant
and 
$ \LL=\TL $ holds on 
$W$.
\end{lemma}

It follows from straightforward calculations that

\begin{eqnarray}
& & \KK \* \es = \sum_{i,j=1, \ldots n} A^{-1}_{ij} \* \ei 
\times 
\bigl\{ f_j \iast g_{1,2} \ast g_{2,3} \ast \phi_s +  \nonumber \\
\label{3odin}
& &  f_j \ast g_{1,2} \iast g_{2,3} \ast \phi_s + 
f_j \ast g_{1,2} \ast g_{2,3} \iast \phi_s 
\bigr\} 
- \ees - \eees, 
\end{eqnarray}
\begin{eqnarray}
& & \KK \* \ees  = 
\sum_{i,j=1, \ldots n} A^{-1}_{ij} \* 
\ei \times \{ f_j \iast g_{1,2} \iast g_{2,3} \ast \phi_s  + \nonumber \\
\label{3dva}
& & f_j \ast g_{1,2} \iast g_{2,3} \iast \phi_s  
\} - \eeees,
\end{eqnarray}
\begin{eqnarray}
\label{3tri}
& & \KK \* \eees  =  \sum_{i,j=1, \ldots n} A^{-1}_{ij} \* 
\ei \* \{ f_j \iast g_{1,2} \ast g_{2,3} \iast \phi_s \}, \\
\label{3chetyre}
& & \KK \* \eeees  =  \sum_{i,j=1, \ldots n} A^{-1}_{ij} \* 
\ei \* \{ f_j \iast g_{1,2} \iast g_{2,3} \iast \phi_s \}.
\end{eqnarray}

Let us introduce the following notations:
\begin{eqnarray}
\label{BCDE}
& & B_{js}=f_j \iast g_{1,2} \ast g_{2,3} \ast \phi_s +
f_j \ast g_{1,2} \iast g_{2,3} \ast \phi_s +
f_j \ast g_{1,2} \ast g_{2,3} \iast \phi_s,
\\
\label{BCDE2}
& & C_{js}=f_j \iast g_{1,2} \iast g_{2,3} \ast \phi_s +
f_j \ast g_{1,2} \iast g_{2,3} \iast \phi_s
,\\
\label{BCDE3}
& & D_{js}=f_j \iast g_{1,2} \ast g_{2,3} \iast \phi_s, \\
\label{BCDE4}
& & E_{js}=f_j \iast g_{1,2} \iast g_{2,3} \iast \phi_s, 
\ \ \ j,s=1, \ldots,n.
\end{eqnarray}

Let us denote by $K$ the matrix of the restriction of the operator $\KK$ on the subspace $ W $ in the basis 
$ \left \{\ei, \eei, \eeei, \eeeei, \ i=1, \ldots, n \right \}. $ Then it follows from 
(\ref{3odin}-\ref{3chetyre}) that

$$
K= \left( \begin{array}{cccc}  A^{-1}\* B   & A^{-1} \* C &  A^{-1} \* D & A^{-1}\* E \\
                           -Id  & 0   &  0 & 0 \\
                           -Id  & 0   &  0 & 0 \\
                             0  & -Id &  0 & 0.
\end{array} \right),
$$
and
$$
Id-K= \left( \begin{array}{cccc} Id- A^{-1}\* B & -A^{-1} \* C   & -A^{-1} \* D & -A^{-1}\* E \\
                                Id  & Id &  0 & 0 \\
                                Id  & 0  & Id & 0 \\
                                 0  & Id &  0 & Id
\end{array} \right),
$$
The matrix $ (Id -K) $ can be easily inverted. The following simple lemma holds.
\newcommand{\Opr}{ (K_1+ K_4 - K_2 - K_3)^{-1}}

\begin{lemma}
Suppose that a matrix $K$  has a $ 4 \times 4$ block form
$$
\left( \begin{array}{cccc}   K_{1}  &  K_{2} &  K_{3} & K_{4} \\
                                 Id  &   Id    &  0      & 0 \\
                                 Id  &     0   &  Id     & 0 \\
                                  0  &    Id   &  0      & Id.
\end{array} \right),
$$
and the matrix $ K_1 +K_4 - K_2 - K_3 $ is invertible.
Then $ \ K^{-1} \ $ is equal to
$$
\left( \begin{array}{cccc}        Q  &  Q\*(K_4-K_2) & -Q\* K_{3}             & -Q \* K_{4} \\
                                 -Q  &  Q\*(K_1-K_3) &  Q\* K_{3}             & Q \* K_{4} \\
                                 -Q  &  Q\*(K_2-K_4) & Id + Q\* K_3           & Q \* K_{4} \\
                                  Q  &  Q\*(K_3-K_1) & -Q\* K_3               & Id - Q \* K_4,
\end{array} \right).
$$
\end{lemma}
where $ \ Q= (K_1 +K_4 - K_2 - K_3)^{-1}. $

In particular for $ K\*(Id -K)^{-1}= (Id-K)^{-1} -Id $ one has 
the following formula:

\begin{equation}
\label{ResK3}
K(Id-K)^{-1}= \left( \begin{array}{cccc}  (A^c)^{-1}\* (A-A^c)   & -(A^c)^{-1} \* (E-C) & (A^c)^{-1}\* D & 
(A^c)^{-1} \* E \\
                                         -(A^c)^{-1}\*A  & (A^c)^{-1} \* (E-C)  & -(A^c)^{-1} \*D  & 
-(A^c)^{-1} \* E \\
                                         -(A^c)^{-1}\* A   &  (A^c)^{-1} \* (E-C) & -(A^c)^{-1}\* D & 
-(A^c)^{-1} \* E \\
                                          (A^c)^{-1}\*A  & -(A^c)^{-1} \* (E-C) -Id  & (A^c)^{-1} \*D  & 
(A^c)^{-1} \* E
\end{array} \right).
\end{equation}
In the calculations above we used the identity
$ (K_1 +K_4 -K_2 -K_3)^{-1}= (Id - A^{-1} \* (B+E -C -D))^{-1}= (Id - A^{-1} \* (A-A^c))^{-1}= (A^c)^{-1} \* A.$

To calculate the values of $ \TL $ on the basis vectors of $W$ we first note that
$ \left(\begin{array}{c} g_{1,2}\ast_c g_{2,3} \ast_c \phi_i \\ g_{2,3} \ast_c \phi_i \\ \phi_i \end{array}
\right) $ can be written as 
$$ \left(\ei - \eei- \eeei + \eeeei\right). $$
Then
\begin{eqnarray}
\label{odinTL3}
& & \TL \* \es  =  
\sum_{i,j=1, \ldots n} (A^c)^{-1}_{ij} \times \\ 
& & \left(\ei - \eei- \eeei + \eeeei\right)  \nonumber \\
& \times & \bigl\{ f_j \iast g_{1,2} \ast g_{2,3} \ast \phi_s + f_j \ast_c g_{1,2}
\iast g_{2,3} \ast \phi_s  + f_j \ast_c g_{1,2} \ast_c g_{2,3} \iast \phi_s \bigr\} \nonumber\\
&-& \ees - \eees + \eeees, \nonumber 
\end{eqnarray}
\begin{eqnarray}
\label{dvaTL3}
& & \TL \* \ees  = 
 \sum_{i,j=1, \ldots n} (A^c)^{-1}_{ij} \times \\ 
& & \left(\ei - \eei- \eeei + \eeeei \right) \nonumber \\
& \times & \bigl\{ f_j \iast g_{1,2} \iast g_{2,3} \ast \phi_s 
+ f_j \ast_c g_{1,2} \iast g_{2,3} \iast \phi_s \bigr\} 
 -  \eeees,  \nonumber
\end{eqnarray}
\begin{eqnarray}
\label{triTL3}
& & \TL \* \eees  = \\
& &  \sum_{i,j=1, \ldots n} (A^c)^{-1}_{ij} \* 
\left(\ei - \eei -\eeei + \eeeei \right) \nonumber \\
&\times& \bigl\{ f_j \iast g_{1,2} \ast g_{2,3} \iast \phi_s \bigr\}, \nonumber 
\end{eqnarray}
\begin{eqnarray}
\label{chetyreTL3}
& & \TL \* \eeees  =  \\
& & \sum_{i,j=1, \ldots n} (A^c)^{-1}_{ij} \* 
\left(\ei - \eei -\eeei + \eeeei \right) \nonumber \\
& \times &  \bigl\{ f_j \iast g_{1,2} \iast g_{2,3} \iast \phi_s \bigr\}. \nonumber
\end{eqnarray}
It follows from (\ref{ResK3}), (\ref{BCDE} - \ref{BCDE4}) and (\ref{odinTL3} -\ref{chetyreTL3}) that
$ \LL = \TL $ on $W$. Lemma is proven.

To finish the proof of Theorem for the case of three classes of particles, we need to
prove the relation $ \TL= \LL$ also on a complement subspace of $ W $ in $ L^2(\mathcal{I}).$
Let us introduce the following subspaces $ V_1 \subset L^2(I_1), \ \ V_2 \subset L^2(I_2), \ \ 
V_3 \subset L^2(I_3) : $
\begin{eqnarray*}
& & V_1:= Span \{ g_{1,2} \iast g_{2,3} \iast \phi_i, \ \ i=1, \ldots,n \}, \\
& & V_2:= Span \{ g_{2,3} \iast \phi_i, \ \ i=1, \ldots, n\},  \\
& & V_3:= Span \{ \phi_i, \ \ i=1, \ldots, n \}. 
\end{eqnarray*}
We already showed that 
\begin{equation}
\label{one}
 \LL \* \left( \begin{array}{c}  g_{1,2} \iast g_{2,3} \iast \phi_i \\ 0 \\ 0 
\end{array} \right) =
\TL \* \left( \begin{array}{c} g_{1,2} \iast g_{2,3} \iast \phi_i \\ 0 \\ 0 \end{array} \right),
\ \ i=1, \ldots,n. 
\end{equation}
Below we will also prove that
\begin{equation}
\label{two}
\LL \* \left( \begin{array}{c}  0 \\ g_{2,3} \iast \phi_i \\ 0 
\end{array} \right) =
\TL \* \left( \begin{array}{c} 0 \\ g_{2,3} \iast \phi_i \\ 0 \end{array} \right),
\ \ i=1, \ldots,n, 
\end{equation}
and
\begin{equation}
\label{three}
\LL \* \left( \begin{array}{c}  0 \\  0 \\ \phi_i 
\end{array} \right) =
\TL \* \left( \begin{array}{c} 0 \\  0 \\  \phi_i \end{array} \right),
\ \ i=1, \ldots,n. 
\end{equation}
Once this is accomplished, it will be enough to prove $ \TL=\LL $ on the subspaces

$ \left( \begin{array}{c}
(V_1)^{\bot} \\ 0 \\0 \end{array} \right), \ \ 
\left( \begin{array}{c} 0\\
(V_2)^{\bot} \\ 0 \end{array} \right), $ and 
$ \left( \begin{array}{c} 0\\
0 \\ (V_3)^{\bot}  \end{array} \right) $ 
of the Hilbert space $ L^2(\mathcal{I})$.
What is more, the 
inveribility of the matrices
$ A^{\mathcal{I}}, \ \ A^{\mathcal{I},1}, \ \ A^{\mathcal{I},2} \ $ implies  that it will be enough to prove
the desired relation  on the subspaces
$ \left( \begin{array}{c}
(V_4)^{\bot} \\ 0 \\0 \end{array} \right), \ \ 
\left( \begin{array}{c} 0\\
(V_5)^{\bot} \\ 0 \end{array} \right) $ and
$ \left( \begin{array}{c} 0\\
0 \\ (V_6)^{\bot} \end{array} \right), $
where
$ V_4= Span \{ \overline{f_j}, j=1, \ldots, n \} \subset L^2(I_1), 
\ \  V_5= Span \{\overline{f_j \iast g_{1,2}}, 
\ j=1, \ldots, n\} \subset L^2(I_2) \ $ and
$V_6= Span \{\overline{f_j \iast g_{1,2} \iast g_{2,3}}, 
\ j=1, \ldots, n\} \subset L^2(I_3). \ \ $ 
Indeed, $ A^{\mathcal{I}}$ 
is the matrix of the scalar products of the basis vectors in 
$V_1 ( $ resp. $ \ V_2, \ V_3)$ and the basis vectors of $V_4 ( $ resp. $ \ V_5, \ V_6)$ in 
$ L^2(I_1) \ ( $ resp. $ L^2(I_2), \ L^2(I_3))$). Therefore,  invertibility of $ A^{\mathcal{I}}$  implies that
the sum of $V_1$ and $(V_4)^{\bot}$ is the whole $L^2(I_1)$ (similarly for $V_2$ and $(V_5)^{\bot}, \ \ 
V_3$ and $(V_6)^{\bot}$). 
We claim the following lemma is true.

\begin{lemma}
$\LL = \TL $ holds on
$ \left( \begin{array}{c}
(V_4)^{\bot} \\ 0 \\0 \end{array} \right), \ \ 
\left( \begin{array}{c} 0\\
(V_5)^{\bot} \\ 0 \end{array} \right) $ and
$ \left( \begin{array}{c} 0\\
0 \\ (V_6)^{\bot} \end{array} \right) $
and  on \\
$ \left( \begin{array}{c} 0 \\ g_{2,3} \iast \phi_i \\ 0 \end{array} \right), \ \  
\left( \begin{array}{c} 0 \\  0 \\  \phi_i \end{array} \right), \ \  i=1, \ldots, n.$
\end{lemma}

The first part is trivial. Indeed,
$ \LL = \TL =0 $ on 
$ \left( \begin{array}{c}
(V_4)^{\bot} \\ 0 \\0 \end{array} \right).$ To prove the second part of the lemma we note that the last identity 
together with (\ref{one})  imply that $ \LL= \TL $ on all 
vectors $ \left( \begin{array}{c} h \\ 0 \\ 0 \end{array} \right). $  Since it is also true that 
$ \LL \eei= \TL \eei, \ \ i=1, \ldots, n, $ we obtain that
$ \LL \* \left( \begin{array}{c} 0 \\ g_{2,3} \iast \phi_i \\ 0 \end{array} \right)=
\TL \* \left( \begin{array}{c} 0 \\ g_{2,3} \iast \phi_i \\ 0 \end{array} \right).$
By the argument presented above, in order to prove
$ \LL = \TL $ on all vectors 
$ \left( \begin{array}{c} 0 \\ h \\ 0 \end{array} \right) $ we need to check this relation on
$ \left( \begin{array}{c} 0\\
(V_5)^{\bot} \\ 0 \end{array} \right). $ 
Let $ f_j \iast g_{1,2} \iast h =0.$  Then
\begin{eqnarray}
\label{TLoho}
& & \TL \  \left( \begin{array}{c}
0\\ h \\0 \end{array} \right) = \sum_{i,j=1}^n (A^c)^{-1}_{ij} \* \bigl\{f_j \ast_c g_{1,2} 
\iast h \bigr\} \times \\
& & \left( \ei - \eei - \eeei + \eeeei \right) \nonumber \\
&-& 
\left( \begin{array}{c}
g_{1,2}\iast h \\ 0 \\0 \end{array} \right). \nonumber
\end{eqnarray}

To calculate $ \LL \  \left( \begin{array}{c}
0\\ h \\0 \end{array} \right) $  we write
\begin{equation}
\label{LLoho}
 \LL \  \left( \begin{array}{c} 0\\ h \\0 \end{array} \right) = 
\KK \  \left( \begin{array}{c} 0\\ h \\0 \end{array} \right) +
\LL \* \KK \  \left( \begin{array}{c} 0\\ h \\0 \end{array} \right) 
\end{equation}
Now, 
\begin{equation}
\label{KKoho}
\KK \  \left( \begin{array}{c} 0\\ h \\0 \end{array} \right) = \sum_{i,j=1}^n A^{-1}_{ij} \*
\bigl\{ f_j \ast g_{1,2} \iast h \bigr\} \* \ei \ - 
\left( \begin{array}{c} g_{1,2} \iast h \\ 0 \\0 \end{array} \right), 
\end{equation}
 and 
\begin{equation}
\label{skoro}
\LL \* \KK \  \left( \begin{array}{c} 0\\ h \\0 \end{array} \right) = 
\sum_{i,j=1}^n A^{-1}_{ij} \*
\bigl\{ f_j \ast g_{1,2} \iast h \bigr\} \ \LL \ei \ 
- \LL \left( \begin{array}{c} g_{1,2} \iast h \\ 0 \\0 \end{array} \right)  
\end{equation}
Since $ f_j \iast g_{1,2} \iast h =0 $ we can claim
$ \LL \left( \begin{array}{c} g_{1,2} \iast h \\ 0 \\0 \end{array} \right)=0.$
Substituting (\ref{odinTL3}) into (\ref{skoro}), and combining (\ref{TLoho}- \ref{skoro})  we arrive at
$\LL \left( \begin{array}{c} 0\\ h \\0 \end{array} \right) = \TL \left( \begin{array}{c} 0\\ h \\0 \end{array}
\right) . \ $ 

So far we have established that  $ \LL = \TL $ holds on all vectors of the form 
$ \left( \begin{array}{c}
h_1 \\ h_2 \\0 \end{array} \right). $  We also proved $ \LL \* \ei = \TL \* \ei, \ \ i=1, \ldots, n, $ (since $
\ei \in W $). Therefore we conclude that $ \LL = \TL $ holds on the vectors
$ \left( \begin{array}{c} 0 \\  0 \\  \phi_i \end{array} \right), \ \  i=1, \ldots, n.$
The last step in the  proof is to show that
$\LL = \TL $ holds on
$ \left( \begin{array}{c} 0\\
0 \\ (V_6)^{\bot} \end{array} \right).  \ \ $
Let $ f_j \iast g_{1,2} \iast g_{2,3} \iast h =0.\ $  Then
\begin{eqnarray}
\label{TLohoho}
& & \TL \  \left( \begin{array}{c}
0\\ 0 \\ h \end{array} \right) = \sum_{i,j=1}^n (A^c)^{-1}_{ij} \* \bigl\{f_j \ast_c g_{1,2} 
\ast_c g_{2,3} \iast h \bigr\} \times \\
& & \left( \ei - \eei - \eeei + \eeeei \right) \nonumber \\
&-& 
\left( \begin{array}{c}
g_{1,2}\ast_c g_{2,3} \iast h \\ 0 \\0 \end{array} \right) - 
\left( \begin{array}{c}
0 \\ g_{2,3} \iast h \\ 0  \end{array} \right). 
\nonumber
\end{eqnarray}
Now,
$ \LL \* \left( \begin{array}{c} 0\\ 0 \\ h \end{array} \right) = 
\KK \* \left( \begin{array}{c} 0\\ 0 \\ h \end{array} \right) +
\LL \* \KK \* \left( \begin{array}{c} 0\\ 0 \\ h \end{array} \right), $ and
\begin{equation}
\label{KKohoho}
\KK \  \left( \begin{array}{c}
0\\ 0 \\ h \end{array} \right) = \sum_{i,j=1}^n A^{-1}_{ij} \* \bigl\{f_j \ast g_{1,2} 
\ast g_{2,3} \iast h \bigr\} \times 
\ei 
-
\left( \begin{array}{c}
g_{1,2}\ast g_{2,3} \iast h \\ 0 \\0 \end{array} \right) - 
\left( \begin{array}{c}
0 \\ g_{2,3} \iast h \\ 0  \end{array} \right). 
\nonumber
\end{equation}
It follows from (\ref{KKohoho}) and previous calculations that  $ \LL = \TL $ holds on the vectors 
$\KK \* \left( \begin{array}{c} 0\\ 0 \\ h \end{array} \right).$ Therefore
\begin{eqnarray*}
& &  \LL \* \left( \begin{array}{c} 0\\ 0 \\ h \end{array} \right) = 
\KK \* \left( \begin{array}{c} 0\\ 0 \\ h \end{array} \right) +
\sum_{i,j=1}^n A^{-1}_{ij} \* \bigl\{f_j \ast g_{1,2} 
\ast g_{2,3} \iast h \bigr\} \times \TL \* \ei \\
& - & \TL \* \left( \begin{array}{c}
g_{1,2}\ast g_{2,3} \iast h \\ 0 \\0 \end{array} \right) - 
\TL \* \left( \begin{array}{c}
0 \\ g_{2,3} \iast h \\ 0  \end{array} \right) .
\end{eqnarray*}
After simple algebraical calculations we obtain the desired identity. The details are left to the reader.
Lemma is proven.
This finishes the proof  of the Theorem for $ M=3.$

\section{General $ M$  Case}

The plan of the proof is the same as in the special case discussed above.
We introduce a $ 2^{M-1} \times n $- dimensional subspace of $ L^2(\mathcal{I}),$ which is invariant under
$\KK$ and $\TL.$ As before we will denote the subspace by $W.$ The basis vectors of $W$ can be divided into
$ 2^{M-1} $ different groups, each group consisting of $n $ elements. The vectors in each group
will be indexed by $ i=1,\ldots, n$. These $2^{M-1}$ different groups can be put in one-to 
one correspondence with the subsets of $\{1,2,3,\ldots, M-1\}.$ 
The empty set will correspond to  vectors denoted by $ e^0_i, \ \ i=1, \ldots, n, $ where
$ e^{(0)}_i $ can be thought as  a $M$- column 
$ e^{(0)}_i = ( g_{1,2} \ast g_{2,3} \ast \ldots \ast g_{M-1,M} \ast \phi_i, \ \ \ 
g_{2,3}\ast \ldots \ast g_{M-1,M}
\ast \phi_i, \ \ \ g_{3,4} \ast \cdots \ast \phi_i, \  \ldots, \ \ \phi_i)^t.$ In the case $M=3$ the 
$2^{3-1} \times n =4 \times n $-dimensional subspace $W$ was 
introduced in the previous section. In particular, for $M=3$ we have $e^{(0)}_i= \ei.$ 
The remaining $2^{M-1} -1$ groups will be indexed by 
$(l_1, \ldots, l_r),$ where $ r=1,2,3, \ldots, \ \ l_1 \geq 1,
\ldots, l_r \geq 1,\ \ \ l_1 + \cdots + l_r \leq M-1.$ 
Each such $r$-tuple corresponds to an $r-element$ subset  $\{ l_1, l_1+l_2,\ldots, l_1+\cdots + \l_r \}$ of
$ \{1,2,3,\ldots,M-1\}.$
The corresponding basis vectors of $W$ will be denoted by  $e^{(l_1,\ldots, l_r)}_i.$ The vector 
$e^{(l_1,\ldots, l_r)}_i$ 
looks
similar to  $e^{(0)}_i$ defined above. It can be again viewed as an $M$-column, but now  only the first
$M- l_1-l_2 -\cdots - l_r$ components are non-zero.  The first component is
$ g_{1,2} \ast \cdots \ast g_{l_1, l_1+1} \iast g_{l_1+1, l_1+2} \ast \cdots \ast
g_{l_1+l_2, l_1+l_2+1} \iast g_{l_1+l_2+1, l_1+l_2+2} \ast \cdots  \phi_i.$ 
The difference with the first component of $e^{(0)}$ is that
after $g_{l_1,l_1+1}, \ g_{l_1+l_2, l_1+l_2+1}, $ etc, (altogether in $r$ places)
the convolution symbol $\ast$ has been replaced by the convolution symbol
$\iast.$ The second component of $e^{(l_1, \ldots,l_r)}$ is
$ g_{2,3} \ast \cdots \ast g_{l_1+1, l_1+2} \iast g_{l_1+2, l_1+3} \ast \cdots
\ast g_{l_1+l_2+1, l_1+l_2+2} \iast g_{l_1+l_2+2, l_1+l_2+2} \ast \cdots   \phi_i.$ 
The difference with the second component of $e^{(0)}$ is that
after $g_{l_1+1,l_1+2}, \ g_{l_1+l_2+1, l_1+l_2+2}, $ etc, (altogether in $r$ places)
the convolution symbol $\ast$ has been replaced by the convolution symbol $\iast.$  
Please note that in  the second component the places where 
symbols $\ast$ has been replaced
by $\iast$ are shifted by 1 in comparison with the first component. The third component is 
$ g_{3,4} \ast \cdots \ast g_{l_1+2, l_1+3} \iast g_{l_1+3, l_1+4} \ast \cdots 
\ast g_{l_2+2, l_2+3} \iast \cdots \phi_i.$ Again, the places where the convolution symbols $\iast$ appear 
have been
shifted by $1$ in comparison with the second component.
In a similar fashion we construct the first $ M- l_1 -l_2 - \cdots - l_r $ components. All  these components
have exactlly $r$ convolution symbols $\iast$ (in the $ (M-l_1-l_2 - \cdots - l_r)$-th component 
the last 
convolution symbol $\iast$ appeares in front of $\phi_i$).
We put the  remaining  $ l_1 + l_2 \cdots + l_r $ components to be zero (please note that they can not be 
constructed
according to the  principle described above if we want to keep the number of convolution symbols 
$\iast$ equal to $r$).

Since the constructions of such nature could be better understood after playing with some examples, we discuss
two of them below.

{\bf Example}

{\it $e^{(1)}_i= (g_{1,2}\iast g_{2,3} \ast g_{3,4} \cdots \ast \phi_i, \ \ g_{2,3} \iast g_{3,4} \ast g_{4,5} 
\cdots
\ast \phi_i, \ 
\ldots, \ \ g_{M-1,M}\iast \phi_i, \ \ 0)^t.$ The last 
component of $e^{(1)}_i$ is zero (and not, say, $\phi_i$)  because we insist that all non-zero components must 
have
the same number of $\iast$ convolutions (in this example the number of $\iast$ convolutions is one;  the
place of  the $ \iast$  convolution is shifted to the right 
by a unit each 
time we go from $k$-th to $k+1$-th component, $ k=1,2,\ldots, M-2.$)}

{\bf Example}

{\it In the case $M=3$ (see section 2)  the basis vectors of $W$ are
\begin{eqnarray*}
& & e^{(0)}_i= \ei, \ \ e^{(1)}_i= \eei, \ \ e^{(2)}_i= \eeei, \\ 
& & e^{(1,1)}_i= \eeeei, \ \ \ 1\leq i \leq n.
\end{eqnarray*}}

\newcommand{\ellri}{e^{(l_1, \ldots, l_r)}_i}
\newcommand{\ellrs}{e^{(l_1, \ldots, l_r)}_s}
\newcommand{\elllri}{e^{(l,l_1, \ldots, l_r)}_i}
\newcommand{\elllrs}{e^{(l,l_1, \ldots, l_r)}_s}
\newcommand{\eclllrs}{e^{(l,l_1, \ldots, l_r),c}_s}
\newcommand{\eclllri}{e^{(l,l_1, \ldots, l_r),c}_i}
We start by calculating the matrix of the restriction of $\KK$ on $W$.
It follows immediately from (\ref{K}) that
\begin{equation}
\label{Kl1lr}
\KK \* \ellrs = \sum_{i=1, \ldots, n} (A^{-1}\* B^{(l_1, \ldots, l_r)})_{is} \* e^{(0)}_i - \sum_{l>0}
\elllrs,
\end{equation}
where
\begin{eqnarray}
\label{Bl1lr}
& & B^{(l_1, \ldots, l_r)}_{ij}= f_i\iast g_{1,2} \ast \cdots 
\ast g_{l_1, l_1+1} \iast g_{l_1+1,l_1+2} \ast \cdots  \phi_j  + \\ 
& & f_i \ast g_{1,2} \iast g_{2,3} \ast \cdots g_{l_1+1, l_1+2}\iast g_{l_1+2, l_1+3} \ast 
\cdots \phi_j + \nonumber \\  
& &  f_i \ast g_{1,2} \ast g_{2,3} \iast g_{3,4} \ast \cdots g_{l_1+2, l_1+3}\iast g_{l_1+3, l_1+4} \ast 
\cdots \phi_j  + \ldots, \nonumber
\end{eqnarray}
where in the first term of the r.h.s. of (\ref{Bl1lr}) 
the convolution symbols $ \iast $ appear after $f_i, \ \ g_{l_1,l_2}, \ \ 
g_{l_1+l_2, l_1+l_2+1}, \ldots,$ (the other convolution symbols are $\ast$), in the second term the convolution 
symbols $\iast$ appear after $g_{1,2}, \ \ g_{l_1+1, l_1+2}, \ \ g_{l_1+l_2+1, l_1+l_2+2}, \ldots,$ etc. 
Altogether, there are $M- l_1 - l_2 - \cdots - l_r $ (and not $M \ $) terms in the sum, because we require
(as before) that each term has the same number of $\iast$ convolutions. 
For $ l+l_1+\cdots + l_r \geq M $ we agree to set $ e^{(l,l_1, \ldots, l_r)}_s =0.$
The same rules apply to  similar notations introduced below. 
Let the kernel $\TL$ be defined by the right hand side of (\ref{Lform}).
Then

\begin{equation}
\label{TLl1lr}
\TL \* \ellrs = \sum_{i=1, \ldots, n} ((A^c)^{-1}\* B^{(l_1, \ldots, l_r),c})_{is} \* e^{(0),c}_i - \sum_{l>0}
\eclllrs,
\end{equation}
where $ e^{(l_1, \ldots, l_r),c}_i  $ is defined after the example below and
$B^{(l_1, \ldots, l_r),c}_{ij}$ is defined in a similar way to $B^{(l_1, \ldots, l_r)}_{ij},$  but with 
a twist. Namely,
\begin{eqnarray}
\label{Bcl1lr}
B^{(l_1, \ldots, l_r),c}_{ij}&=& f_i\iast g_{1,2} \ast \cdots g_{l_1, l_1+1} \iast g_{l_1+1, l_1+2} \ast
\cdots \ast \phi_j  \\
& + &
f_i \ast_c g_{1,2} \iast g_{2,3} \ast \cdots g_{l_1+1, l_1+2}\iast g_{l_1+2, l_1+3} \ast 
\cdots \phi_j  \nonumber \\
& + & \ f_j \ast_c g_{1,2} \ast_c g_{2,3} \iast g_{3,4} \ast \cdots g_{l_1+2, l_1+3} \iast 
g_{l_1+3, l_1+4} \ast \cdots \phi_j \nonumber \\
& +&  \ \ldots. \nonumber
\end{eqnarray}
The first term of the sum  (\ref{Bcl1lr}) is the same as the first term of the sum (\ref{Bl1lr}).
The only difference between the second term in (\ref{Bl1lr}) and the second term in 
(\ref{Bcl1lr}) is that in the second term of (\ref{Bcl1lr})  the first convolution symbol (between $ f_i$ 
and $g_{1,2}$) is $ \ast_c,$ and not 
$\ast$. In the third term of (\ref{Bcl1lr}) 
the first two convolution symbols are $\ast_c$ and the other are the same as in 
the third term of (\ref{Bl1lr}), etc.

{\bf Example}

{\it Verify that $B^{(0),c} = A - A^c.$}

To make sense of (\ref{TLl1lr}) we also have to define $e^{(0),c}_i, $ and, in general, 
$ e^{(l_1, \ldots, l_r),c}_i $. We write
\begin{equation}
\label{ec0i}
e^{(0),c}_i = ( g_{1,2} \ast_c \ldots \ast_c g_{M-1,M} \ast_c \phi_i, 
g_{2,3}\ast_c \ldots \ast_c g_{M-1,M}
\ast_c \phi_i, \ldots, \phi_i)^t.
\end{equation}
In other words the difference between $ e^{(0),c}_i \ $ and $ e^{(0)}_i $ is that in 
$ e^{(0),c}_i \ $ all convolution symbols $\ast$ are replaced by  the convolution symbols $\ast_c$.
To obtain $e^{(l_1,\ldots,l_r),c} \ $ from
$e^{(l_1,\ldots,l_r)} \ $ we have to replace  
in each component of $e^{(l_1,\ldots,l_r),c} \ $  
the first (from the left) $ l_1 -1$  
convolution symbols $\ast$ by $\ast_c$ (in other words we do it untill we meet the first symbol 
$\iast,$ at which point we 
stop).

The inclusion-exclusion principle implies:
\begin{equation}
\label{inex1}
e^{(0),c}_i = e^{(0)}_i + \sum_{r =1,2,\ldots} \ \sum_{l_1, \ldots, l_r} \* (-1)^r \* \ellri,
\end{equation}
where the summation is over all possible $1 \leq l_1, \ldots,  l_r, \ \ l_1+ \cdots + l_r \leq M-1.$
Similarly,
\begin{equation}
\label{inex2}
e^{(l_1, \ldots, l_r),c}_i = e^{(l_1, \ldots, l_r)}_i + \sum_{p =1,2,\ldots, l_1-1} \ \sum_{k_1, \ldots, k_p} 
\* (-1)^p \* e^{(k_1,\ldots, k_p,  t,  l_1, \ldots, l_r)},
\end{equation}
where $ t= l_1 -k_1 -\cdots - k_p,$ and
the summation is defined over all possible $ k_1, \ldots, k_p,$ such that $ 1 \leq k_1, \ldots,  k_p, \ \ 
k_1 + \cdots + k_p < l_1.$ In particular,
$e^{(l_1, \ldots, l_r),c}_i = e^{(l_1, \ldots, l_r)}_i $ for $ l_1=1.$

\begin{lemma}
The operators $ \KK, \ \ \TL \ $ leave $W$ invariant and $ \LL = \TL $ holds on $W.$
\end{lemma}

We have to show that $ (Id + \TL)\* (Id - \KK) = Id$ on $ W.$
By linearity it is enough to check the identity on the basis vectors.
It follows from (\ref{Kl1lr}) and (\ref{TLl1lr}) that
\begin{equation}
\label{IdKl1lr}
(Id - \KK) \* \ellrs = - \sum_{i=1, \ldots, n} (A^{-1}\* B^{(l_1, \ldots, l_r)})_{is} \* e^{(0)}_i + 
\sum_{l> 0}
\elllrs + \ellrs,
\end{equation}
and
\begin{eqnarray}
\label{IdLIdK}
& & (Id + \TL) \* (Id - \KK) \* \ellrs =  - \sum_{i=1}^n 
(A^{-1}\* B^{(l_1, \ldots, l_r)})_{is} \* e^{(0)}_i + \sum_{l> 0} \elllrs  \ + \ellrs \nonumber \\
& - & \sum_{j=1}^n \ \left( (A^c)^{-1} \* B^{(0),c} \* A^{-1} \* B^{(l_1,\ldots, l_r)} \right)_{js} \ 
e^{(0),c}_j 
\ + \ \sum_{i=1}^n  \ (A^{-1} \* B^{(l_1,\ldots, l_r)})_{is} \ \times \ \sum_{p\geq 1} e^{(p),c}_i \nonumber
\\
&+ & \sum_{j=1}^n \sum_{l> 0} \ \left( (A^c)^{-1} \* B^{(l,l_1, \ldots, l_r),c} \right)_{js} \* e^{(0),c}_j \ - 
\sum_{p>0, l>0} e^{(p,l,l_1,\ldots, l_r),c}_s \nonumber \\
&+ &  \sum_{j=1}^n \left( (A^c)^{-1} \* B^{(l_1, \ldots, l_r),c} \right)_{js} \* e^{(0),c}_j \ - 
\sum_{p>0} e^{(p,l_1,\ldots, l_r),c}_s . 
\end{eqnarray}
Using the identity $ B^{(0),c}=A-A^c $ we can rewrite  (\ref{IdLIdK}) as 
$ (Id + \TL) \* (Id - \KK) \* \ellrs =  S_1 + S_2 +S_3, $ where
\begin{eqnarray}
\label{SSS}
& & S_1= \sum_{i=1}^n (A^{-1}\* B^{(l_1, \ldots, l_r)})_{is} \* \bigl( - e^{(0)}_i +
e^{(0),c}_i +
\sum_{p=1}^{M-1} e^{(p),c}_i \bigr),  \\
& & S_2=   \sum_{j=1}^n \ \left( - (A^c)^{-1} \* B^{(l_1,\ldots, l_r)} 
+ (A^c)^{-1} \* B^{(l_1, \ldots, l_r),c} 
+ \sum_{l> 0} \  (A^c)^{-1} \* B^{(l,l_1, \ldots, l_r),c}  
\right)_{js} \ e^{(0),c}_j  \\
& & S_3= \sum_{l> 0} \elllrs  \ + \ellrs \ - \ \sum_{p>0, l>0} e^{(p,l,l_1,\ldots, l_r),c}_s \ - \ 
\sum_{p>0} e^{(p,l_1,\ldots, l_r),c}_s . 
\end{eqnarray}
We claim that $ S_1= S_2=0 \ $  and $ \ S_3 = \ellrs.$
Indeed, an easy inductive argument (similar to the one showing that $B^{(0),c}=A-A^c $) gives
\begin{eqnarray}
\label{identities}
& & \sum_{p=1}^{M-1} e^{(p),c}_i = e^{(0)}_i - e^{(0),c}_i, \\
& & \sum_{l> 0} \  B^{(l,l_1, \ldots, l_r),c}  = B^{(l_1,\ldots, l_r)} - B^{(l_1, \ldots, l_r),c}.
\end{eqnarray}
To tackle $ S_3$ we note that
\begin{eqnarray}
\label{ravenstvo}
& & \elllrs - e^{(l,l_1,\ldots, l_r),c}= \sum_{1 \leq p \leq l-1} e^{(p,l-p,l_1,\ldots, l_r),c}_s, \\
& & \sum_{l>0} \sum_{1 \leq p \leq l-1} e^{(p,l-p,l_1,\ldots, l_r),c}_s = 
\sum_{p>0, l>0} e^{(p,l,l_1,\ldots, l_r),c}_s, 
\end{eqnarray}
which immediately implies $S_3= \ellrs.$
Lemma is proven.

To finish the proof of the theorem we need to show that $ \LL = \TL $ on a complement subspace of $W$.
We proceed the same way as in section 3. Let us introduce the following subspaces
$ V_1 \subset L^2(I_1), \ \ V_2 \subset L^2(I_2), \ldots, \ 
V_M \subset L^2(I_M) : $
\begin{eqnarray*}
& & V_1:= Span \{ g_{1,2} \iast g_{2,3} \iast g_{3,4} \iast \cdots \iast \phi_i, \ \ i=1, \ldots,n \}, \\
& & V_2:= Span \{ g_{2,3} \iast g_{3,4} \iast \cdots \iast \phi_i, \ \ i=1, \ldots, n\},  \\
& & ....\\
& & V_M:= Span \{ \phi_i, \ \ i=1, \ldots, n \}. 
\end{eqnarray*}

Consider a vector 
$e^{(1,1,\ldots,1)}_i \ $ (i.e. $ l_1=l_2=\ldots =l_{M-1}=1 \ $) which
has the first component $ g_{1,2} \iast g_{2,3} \iast g_{3,4} \iast \cdots \iast \phi_i, $  and the other 
components zero. We already proved that $ \LL e^{(1,1,\ldots,1)}_i = \TL e^{(1,1,\ldots,1)}_i,$  since
$e^{(1,1,\ldots,1)}_i \  \in W.$ In order to prove $\TL=\LL $ on all vectors of the form $ (h,0,0 \ldots, o)^t$
it is enough to prove the relations for $h$ orthogonal (in $ L^2(I_1) $ ) to 
$g_{1,2} \iast g_{2,3} \iast g_{3,4} \iast \cdots \iast \phi_i. $  The invertibility of matrix 
$ A^{\mathcal{I}} $ implies
that it is enough to prove it for $h$ orthogonal to $\overline{f_j}, \ \ j=1, \ldots, n$,  which is trivial since
both $ \LL $ and $\TL $ are identically zero on such $ (h,0,0,\ldots, 0)^t.$
We proceed now by induction. Suppose that we have already established $ \LL = \TL $ on the subspace
$ L^2(I_1) \bigoplus \cdots \bigoplus L^2(I_{m-1}) \bigoplus \{0\} \cdots \{0\}$ 
( i.e. on the vectors of the form
$(h_1, h_2, \ldots, h_{m-1}, 0,\ldots, 0)^t),$ where $ 2 \leq m \leq M. \ $  
We will deduce that the same identity holds on
$ L^2(I_1) \bigoplus \cdots \bigoplus L^2(I_{m}) \bigoplus \{0\} \cdots \{0\}.$ 
Consider the vector $ e^{(m, 1,1,\ldots,1)}_i, \ \ $ ( i.e. $ \ l_1=m, \ l_2=\ldots = l_{M-m-1}=1 \ $).
Since this vector belongs to $W,$ we have $ \LL \* e^{(m, 1,1,\ldots,1)}_i 
= \TL \* e^{(m, 1,1,\ldots,1)}_i .$ The $m$-th component of this vector is equal to
$ g_{m,m+1}\iast \ldots \iast \phi_i.$ 
The inductive assumption then implies that $ \LL = \TL $ on the vector $ (0,0, \ldots, g_{m,m+1}\iast \ldots
\iast \phi_i, 0 \ldots, 0)^t, \ \ i=1, \ldots,n.$  Using the invertibility of the matrix $A^{\mathcal{I}}$
and arguing as above, we obtain that in order 
to prove $ \LL = \TL $ on $ \{0\} \bigoplus \cdots \bigoplus L^2(I_m) \bigoplus \cdots 
\{0\} $ it is enough to establish the relation only on the vectors $ (0,\ldots, h, \ldots, 0)^t,$ where only $m$-th component is non-zero, 
and $f_i \iast g_{1,2} \iast \cdots \iast g_{m-1,m} \iast h =0.$
We have
\newcommand{\h}{(0,\ldots, h, \ldots, 0)^t}
\begin{eqnarray}
\label{london}
& & \KK  \  \h = \sum_{i,j =1, \ldots, n} A^{-1}_{ij} \* \bigl\{f_j \ast g_{1,2} \ast
\cdots \ast g_{m-1,m}\iast h \bigr\} \* e^{(0)}_i \nonumber \\
& -&  (g_{1,m} \iast h, \ g_{2,m}\iast h, \ldots,
g_{m-1,m}\iast h, 0 \ldots, 0)^t,
\end{eqnarray}
where we remind the reader that $g_{l,m}= g_{l,l+1}\ast \cdots \ast g_{m-1,m}\ $ for $ 1\leq l < m \leq M $.
It follows from (\ref{london}) and the inductive assumption that $ \LL \* \left( \KK \* \h \right) =
\TL \* \left( \KK \* \h \right).$  Therefore
\begin{eqnarray}
\label{moscow}
& & \LL \* \h = \KK \* \h  \ + \ \TL \left( \KK \* \h \right)  \\
& = &
\sum_{i,j =1, \ldots, n} A^{-1}_{ij} \* \bigl\{f_j \ast g_{1,2} \ast
\cdots \ast g_{m-1,m}\iast h \bigr\} \* e^{(0)}_i  \ 
 -  (g_{1,m} \iast h, \ g_{2,m}\iast h, \ldots,
g_{m-1,m}\iast h, 0 \ldots, 0)^t \nonumber \\
&+& \sum_{k,j=1}^n \left((A^c)^{-1} \ (A-A^c) \ A^{-1} \right)_{kj} \  \bigl\{f_j \ast g_{1,2} \ast
\cdots \ast g_{m-1,m}\iast h \bigr\} \times e^{(0),c}_k  \nonumber \\
&-&  
\sum_{i,j =1, \ldots, n} A^{-1}_{ij} \* \bigl\{f_j \ast g_{1,2} \ast
\cdots \ast g_{m-1,m}\iast h \bigr\} \*  \left( \sum_{l\geq 1} \ e^{(l),c}_i   \right) \nonumber \\
& -& \sum_{i,j=1}^n (A^c)^{-1}_{ij} \* \bigl\{ f_j \iast g_{1,m}\iast h \ + \ f_j \ast_c g_{1,2} \iast g_{2,m}
\iast h \ + \ f_j \ast_c g_{1,2} \ast_c g_{2,3}\iast g_{3,m} \iast h \ + \ldots \bigr\} \times e^{(0),c}_i  \nonumber \\
\end{eqnarray}
\begin{eqnarray*}
&+& (g_{1,2}\iast g_{2,m} \iast h, 0,\ldots,o)^t \ + \ (g_{1,2}\ast_c g_{2,3}\iast g_{3,m} \iast h, \ 
g_{2,3} \iast g_{3,m} \iast h, 0 ,\ldots, 0)^t \ +   \nonumber \\
&+& (g_{1,2}\ast_c g_{2,3} \ast_c g_{3,4} \iast g_{4,m} \iast h, \ g_{2,3}\ast_c g_{3,4} \iast g_{4,m} \iast h,
\ g_{3,4} \iast g_{4,m} \iast h, \ 0, \ldots, 0)^t  \ + \ldots  \nonumber\\ 
&+&  \ (g_{1,2}\ast_c g_{2,3} \ast_c \ldots 
\ast_c g_{m-2,m-1} \iast g_{m-1,m} \iast h, \  
\ldots, \ g_{m-2,m-1} \iast g_{m-1,m} \iast h, \ 0, \ldots, \ 0)^t  \nonumber 
\end{eqnarray*}
(the third term of the r.h.s. of (\ref{moscow}) can be simplified since
$ \ \ (A^c)^{-1} \ (A-A^c) \ A^{-1} = (A^c)^{-1} - A^{-1}).\ $
In a similar way 
\begin{eqnarray}
\label{tokyo}
& & \TL  \  \h = \sum_{i,j =1, \ldots, n} (A^c)^{-1}_{ij} \* \bigl\{f_j \ast_c g_{1,2} \ast_c
\cdots \ast_c g_{m-1,m}\iast h \bigr\} \* e^{(0),c}_i \nonumber \\
& -&  (g_{1,2} \ast_c g_{2,3} \ast_c \cdots \ast_c g_{m-1,m} \iast h, \ g_{2,3} \ast_c 
\cdots \ast_c g_{m-1,m}\iast h, \ldots,
g_{m-1,m}\iast h, 0 \ldots, 0)^t.
\end{eqnarray}
To see that right hand sides of (\ref{moscow}) and (\ref{tokyo}) coincide we note that the following three 
identities hold. The first one :
\begin{eqnarray}
\label{sanfran}
& & (g_{1,2}\iast g_{2,m} \iast h, \ 0,\ldots,0)^t \ + \ (g_{1,2}\ast_c g_{2,3}\iast g_{3,m} \iast h, \ 
g_{2,3} \iast g_{3,m} \iast h, \ 0 ,\ldots, 0)^t \  \nonumber \\
&+&  (g_{1,2}\ast_c g_{2,3} \ast_c g_{3,4} \iast g_{4,m} \iast h, 
\ g_{2,3}\ast_c g_{3,4} \iast g_{4,m} \iast h,
\ g_{3,4} \iast g_{4,m} \iast h, \ 0, \ldots, 0)^t  \ + \ldots  \nonumber \\
&+&   (g_{1,2}\ast_c g_{2,3} \ast_c \ldots 
\ast_c g_{m-2,m-1} \iast g_{m-1,m} \iast h, \  
\ldots, \ g_{m-2,m-1} \iast g_{m-1,m} \iast h, \ 0, \ldots, \ 0)^t \nonumber \\
&=& (g_{1,m} \iast h, \ g_{2,m}\iast h, \ldots,
g_{m-1,m}\iast h, 0 \ldots, 0)^t \nonumber \\ 
& -&  (g_{1,2} \ast_c g_{2,3} \ast_c \cdots \ast_c g_{m-1,m} \iast h, \ g_{2,3} \ast_c 
\cdots \ast_c g_{m-1,m}\iast h, \ldots,
g_{m-1,m}\iast h, 0 \ldots, 0)^t.
\end{eqnarray}
The second one :
\begin{eqnarray}
\label{wales}
& & f_j \iast g_{1,m}\iast h \ + \ f_j \ast_c g_{1,2} \iast g_{2,m}
\iast h \ + \ f_j \ast_c g_{1,2} \ast_c g_{2,3}\iast g_{3,m} \iast h \ + \ldots   \\
&+& f_j \ast_c g_{1,2} \ast_c \cdots \ast_c g_{m-2,m-1} \iast g_{m-1,m} \iast h 
= f_j \ast g_{1,m}\iast h 
 - \ f_j \ast_c g_{1,2} \ast_c
\cdots \ast_c g_{m-1,m}\iast h. \nonumber
\end{eqnarray}
And the last one is 
$  \ \sum_{p=1}^{M-1} e^{(p),c}_i = e^{(0)}_i - e^{(0),c}_i, \ \ $ which we already used before. 
The second identity allows us to cancel the terms which have the coefficients $ \ (A^c)^{-1}_{ij}, \ $ the third 
identity
allows us to cancel the terms which have the coefficients $ \ A^{-1}_{ij}, \ $ the first identity allows us to 
cancel the terms which contain no coefficients of the form $ (A^c)^{-1}_{ij}, \ \ A^{-1}_{ij}.$
All three 
identities have simple inductive proofs relying on a telescoping property of the sums in question.  
We show  the proof of (\ref{wales}). The proofs of the other two identities are very similar.
The main (simple) observation is that
$$
f_j \ast_c g_{1,2} \ast_c \cdots \ast_c g_{m-2,m-1} \iast g_{m-1,m} \iast h 
+ \ f_j \ast_c g_{1,2} \ast_c
\cdots \ast_c g_{m-1,m}\iast h = f_j \ast_c g_{1,2} \ast_c \cdots \ast_c g_{m-2,m-1} \ast g_{m-1,m} \iast h.
$$
Taking the sum of the r.h.s. of the last formula and the second to last term of the l.h.s. of  (\ref{wales})
we obtain
$$ f_j \ast_c g_{1,2} \ast_c \cdots \ast_c g_{m-3,m-2} \ast g_{m-2,m-1} \ast g_{m-1,m} \iast h. $$
At the next step we sum the obtained expression with the third to the last term of the l.h.s. of (\ref{wales}),
etc. Repeating the procedure the appropriate number of times we obtain
$ f_j \ast g_{1,2} \ast \cdots \ast g_{m-1,m}\iast h, $ which finishes the proof of (\ref{wales}).
Theorem is proven.

\vspace{0.25cm}
\noindent

We have learned recently from Harnad that he was able to generalize our theorem to dualization with respect to measures modified 
by arbitrary sets of weight functions (\cite{H}).

\vspace{0.25cm}
\noindent

{\bf Acknowledgements.}\, 
It is a pleasure to thank the referees for useful suggestions and John Harnad for letting me know about the preprint \cite{H}

\def\cmp{{\it Commun. Math. Phys.} }

\end{document}